\documentclass{elsarticle}  

\usepackage{lineno,hyperref}
\modulolinenumbers[5]

\usepackage{graphicx}
\usepackage{epstopdf, epsfig}
\usepackage[usenames, dvipsnames]{color}

 \usepackage{url}
\usepackage{hyperref}
\usepackage[capitalise]{cleveref}

\hypersetup{
	colorlinks   = true, 
	urlcolor     = black, 
	linkcolor    = blue, 
	citecolor    = ForestGreen 
}

\usepackage{siunitx}
\usepackage{xfrac}
\usepackage{nicefrac}

\usepackage{amsmath}
\usepackage{amssymb}
\usepackage[labelsep=period]{caption}
\usepackage{subcaption}

\usepackage{natbib}
\usepackage{gensymb}

\graphicspath{ {./graphics/} }
\begin{document}

\begin{frontmatter}

\title{Pressure-driven dynamics of liquid plugs in rectangular microchannels: influence of the transition between quasi-static and dynamic film deposition regimes.}

\author{S. Signe Mamba}
\author{F. Zoueshtiagh}
\author{M. Baudoin \corref{cor1}}

\cortext[cor1]{Corresponding author}
\address{ Univ. Lille, CNRS, Centrale Lille, ISEN, Univ. Valenciennes, UMR 8520 - IEMN, International laboratory LIA/LICS, F-59000 Lille, France}
\ead{michael.baudoin@univ-lille1.fr}

\begin{abstract}
In this paper, we study experimentally and theoretically the dynamics of liquid plugs in rectangular microchannels for both unidirectional and cyclic pressure forcing. In  both cases, it is shown that the transition between quasi-static and dynamic film deposition behind the liquid plug leads to a dramatic acceleration of the plug, rapidly leading to its rupture. This behaviour proper to channels with sharp corners is recovered from a reduced dimension model based on previous theoretical and numerical developments. In addition, it is shown for cyclic periodic forcing that the plug undergoes stable periodic oscillations if it remains in the quasi-static film deposition regime during the first cycle, while otherwise it accelerates cyclically and ruptures. The transition between these two regimes occurs at a pressure-dependent critical initial length, whose value can be predicted theoretically.
\end{abstract}


\end{frontmatter}


\section{Introduction} \label{introduction}

Two-phase gas-liquid flow in microfluidic devices is a hydrodynamic problem with practical applications in a variety of engineered systems including flows in microreactors \citep{song2003microfluidic,gunther2004transport,assmann2011extraction,sobieszuk2012hydrodynamics}, enhanced oil recovery \citep{havre2000taming}, flow in porous media \citep{lenormand1983mechanisms,dias1986network,stark2000motion}, film coating and biomechanical systems as pulmonary flows \citep{kamm1989airway,heil2008mechanics,grotberg2011respiratory}. The significant scope of the topic motivated early studies on the dynamics of these interfacial flows, first in cylindrical tubes \citep{fairbrother1935,taylor1961deposition,bretherton1961} and soon after in polygonal channels \citep{saffman1958penetration,jensen1987effect,ratulowski1989transport,wong1995motion2,kolb1991coating,thulasidas1995bubble,de2007scaling,de2008steady,ijmf_han_2009,ijmf_han_2011}. In the last decades, the interest in segmented gas-liquid flows in polygonal, and in particular, rectangular microchannels has further grown with the development of soft lithography techniques  in microfluidics \citep{duffy1998rapid,quake2000micro}, which enable simple design of complex microchannels with rectangular cross-sections \citep{anderson2000fabrication}. 

A finite volume of liquid (liquid plug or slug) that is displaced by an air finger at constant flow rate or pressure head (Taylor flow) in a tube, leaves on the walls a trailing liquid film. Its thickness can be quantified by the so-called wet fraction, which is the proportion of the tube section occupied by the liquid film. \textit{In cylindrical capillary tubes}, this parameter increases monotonically with the dimensionless velocity of the meniscus (the so-called capillary number Ca) with a $Ca^{2/3}$ law \citep{bretherton1961} at low capillary number. This law can be further extended to larger capillary number as demonstrated by \cite{aussillous2000quick}. In \textit{polygonal microchannels} however, a transition occurs at a critical capillary number between two radically different regimes: Under this critical parameter, the shape of the meniscus remains close to the static one and most of the fluid deposition occurs in the corners of the tube. Indeed, the static shape of the meniscus cannot follow the singular shape of the sharp edges, which leads to significant deposition in the tube corners, even at vanishingly small capillary number. Thus, the wet fraction slightly relies on the plug dynamics and the relative variation of the wet fraction with the capillary number remains weak. \citep{wong1995motion1,wong1995motion2}. Above this critical number, the fluid deposition resulting from the deformation of the rear meniscus induced by the flow overcomes the static one. In this case, the wet fraction becomes again strongly dependent over the capillary number, similarly to what is observed in cylindrical channels. Nevertheless, this process in polygonal channels also depends  on the tube geometry \citep{jensen1987effect,wong1995motion1,wong1995motion2,de2007scaling,de2008steady}. \cite{de2007scaling,de2008steady} showed that it is possible to extend the laws introduced for cylindrical tubes to rectangular tubes, providing the introduction of an aspect-ratio-dependent capillary number . 

This liquid film deposition process induces a dramatic acceleration of a liquid plug when it is pushed at constant pressure head \citep{baudoin2013airway}. Indeed the diminution of the plug size leads to a reduction of the viscous resistance of the plug to motion, itself leading to an acceleration of the plug and thus more fluid deposition. More recently, it has been shown experimentally by \cite{magniez2016dynamics} that the inverse behaviour (progressive slow down and growth of the liquid plug) might also be observed in prewetted capillary tubes depending on the value of the driving pressure and the thickness of the prewetting film. The acceleration and rupture of a liquid plug has also been evidenced in complex tree geometries \citep{baudoin2013airway,song2011air} and for cyclic pressure forcing \citep{jfm_baudoin_2018}. In the latter case however, both the diminution of the viscous resistance and interfacial resistance (due to lubrication effects) at each cycle contribute to the plug acceleration and breaking. 

Nevertheless, all the aforementioned studies were conducted in cylindrical tubes or at capillary numbers well above the critical capillary number. In this paper, we study experimentally and theoretically the influence of the transition between quasi-static and dynamic fluid deposition process on the dynamics of liquid plugs in rectangular channels pushed either with a unidirectional or a cyclic pressure forcing. In both cases, it is shown that the transition between these two regimes leads to a dramatic acceleration of the plug eventually leading to its rupture.  For cyclic forcing, it is shown that under a critical length the plug dynamics is unstable and leads to the plug rupture while above it is stable and periodic; the experimental results are recovered from a reduced dimension model, inspired from previous theoretical developments by \cite{baudoin2013airway}, \cite{magniez2016dynamics} and \cite{jfm_baudoin_2018} adapted here to take into account (i) the modifications of the laws in the rectangular geometry, (ii) lubrication effects resulting from the back and forth motion of the liquid plug on a prewetted tube, and (iii) the transition between quasi-static and dynamic film deposition. The first and second sections provide the experimental and model details. The third and fourth sections explore respectively the response of liquid plugs to unidirectional and periodic cyclic pressure forcings and compare the observed dynamics to results in cylindrical tubes.

\section{Method} \label{Experimental_details}

\begin{figure}
	\centerline{\includegraphics[width=.9\linewidth]{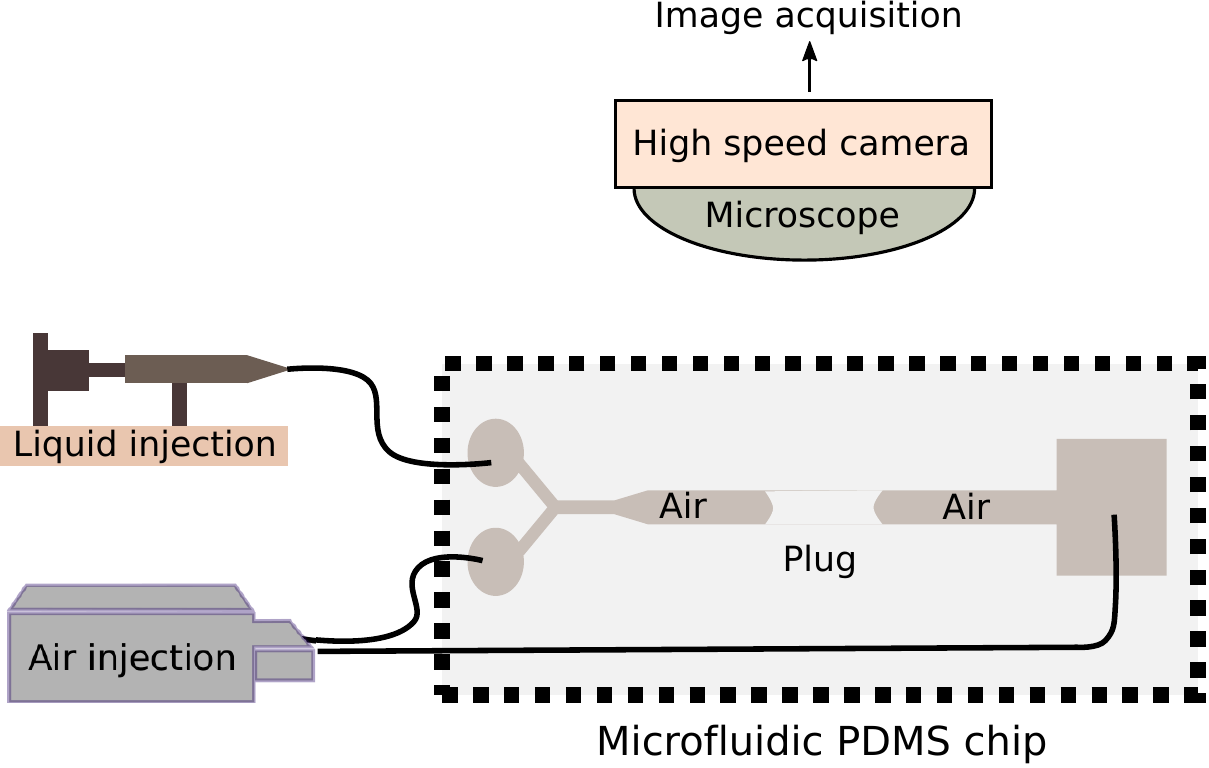}}
	\caption{Sketch of the experimental set-up. A liquid plug is created inside a rectangular PDMS microfluidic channel by pushing alternatively some liquid and some air at a Y-junction. The liquid is pushed by a syringe pump while the air is pushed by a MFCS pressure controller. Then the liquid plug is moved by the pressure controller either in one direction (unidirectional driving) or alternatively in both directions (cyclic periodic driving).}
	\label{setup}
\end{figure}

The experimental setup is represented on \cref{setup}. The experiments are conducted in rectangular  polydimethylsiloxane (PDMS) microfluidic channels of high aspect ratio obtained by standard photolithography techniques: A mold is etched by
depositing a layer of photoresist resin (Microchem, SU8-2035) on a silicon wafer. This layer is spin-coated and patterned by standard
photolithography. The spin-coating speed combined with the choice of the photoresist set the height $h = 45 \pm 2 \mu $m of the microfluidic channels, while the width $w =  785 \mu $m is controlled by the design of the patterned masks which is used during the UV exposure. These values of the channel width and height were measured afterwards with a profilometer (Dektak XTL). After exposure, the film is developed in an organic solvent solution (SU-8 developer) to yield the negative of the channel design. This SU8 mold was used to pour PDMS (Dow Corming, Sylgard 184) whose polymerisation was obtained by curing it at $100^\circ$C. The microfluidic channel is then cut out and bonded on a glass microscope slide by passing the two surfaces in an oxygen plasma. The microscope slides are covered by a thin PDMS membrane in order to guarantee identical boundary conditions to the four channel walls. The total length of the channel is 6 cm.

\begin{figure}
	
	\begin{subfigure}{0.5\textwidth}
		\hspace*{0.4cm} 
		\includegraphics[width=0.9\linewidth, height=4cm]{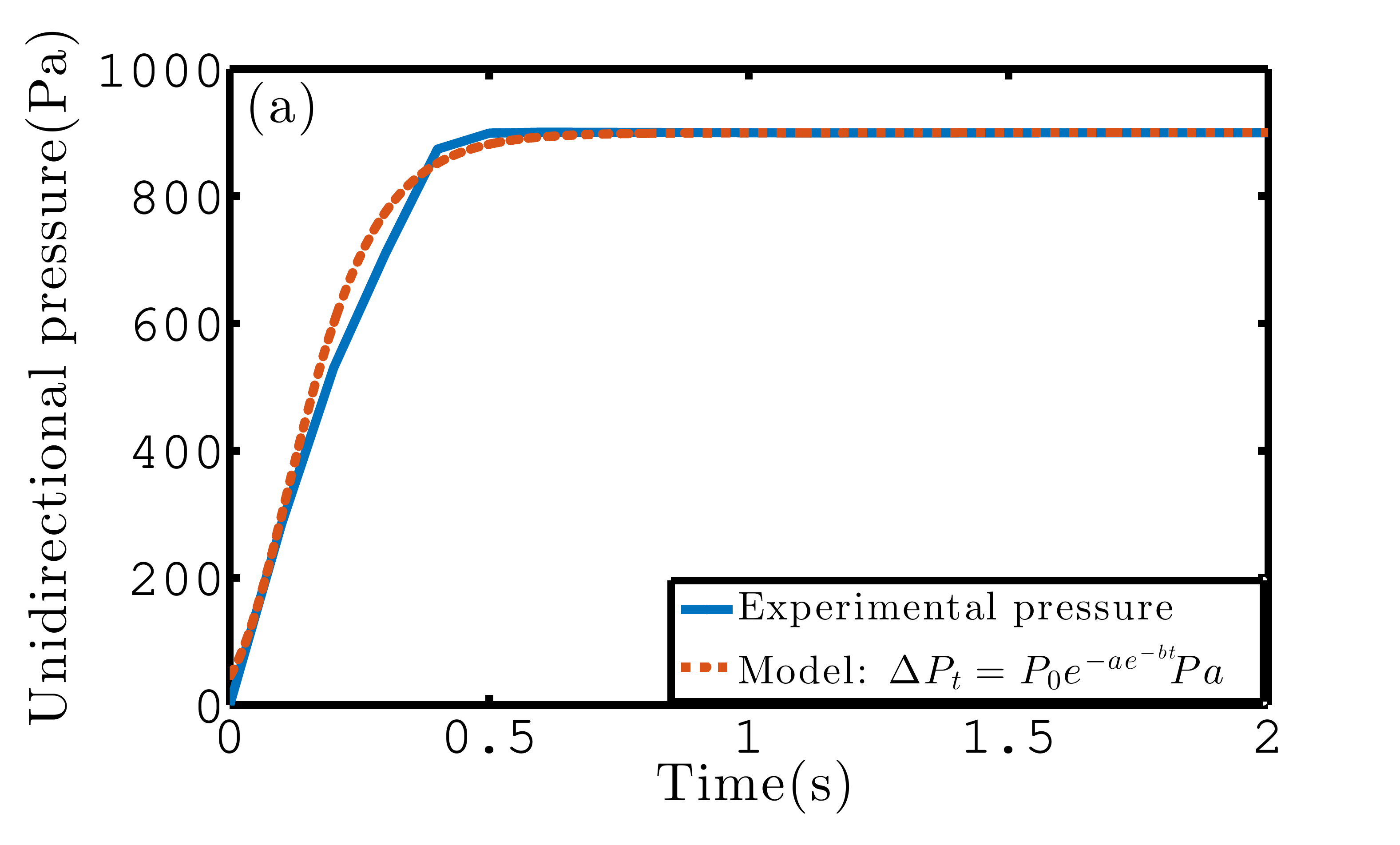} 
	\end{subfigure}
	\begin{subfigure}{0.5\textwidth}
		\hspace*{0.2cm} 
		\includegraphics[width=0.9\linewidth, height=4cm]{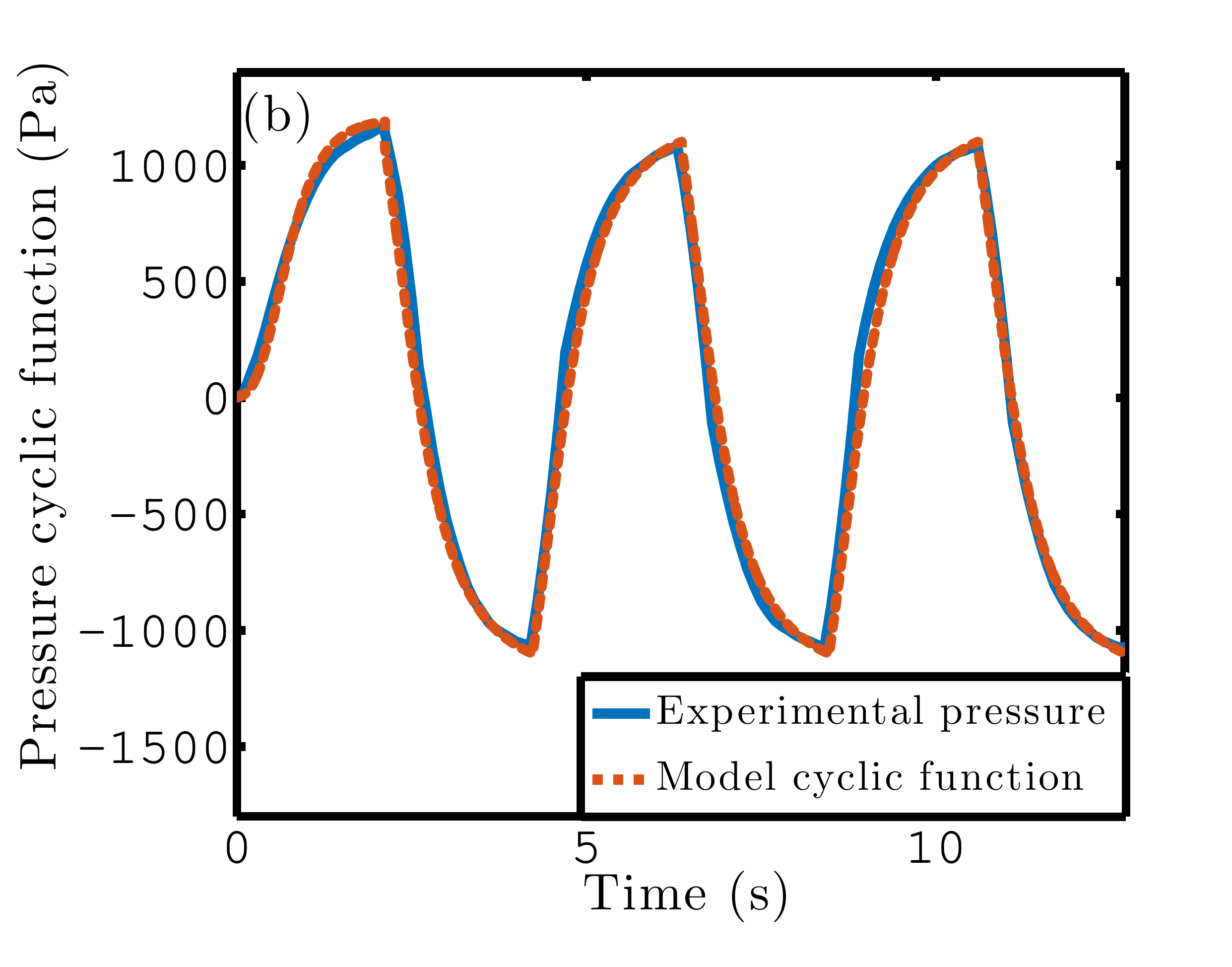}
	\end{subfigure}
	
	\caption{ (a) Unidirectional pressure forcing measured experimentally (blue solid line) and approximated by the formula (red dotted line):  $\Delta P_t = P_0e^{-a e^{-bt}}$ with $P_0=900$ Pa, $a=3$ and the growth rate $b=10$. (b) Cyclic pressure forcing imposed by the MFCS pressure controller measured experimentally (blue solid line) and approximated by the analytical formula  $\Delta P_t = 1200e^{-6 e^{-3t}}$ Pa for $t \in [0;T]$, and $\Delta P_t = {(-1)}^{n} (P_c - P_d)$ for $t \in [nT;(n+1)T]$ with $P_c = 1200 e^{-2.5e^{-3(t-nT)}}$ Pa and $P_d = 1200 e^{-1.2(t-nT)} e^{-0.02e^{-1.2(t-nT)}}$ Pa (red dotted line) and $T=2.12$ is the half period.}
	\label{Pressure_forcing}
\end{figure}

Then, perfluorodecalin (PFD) liquid plugs are created in the channel by pushing alternatively some liquid and some air at a Y-junction with a syringe pump and a MFCS Fluigent pressure controller respectively, connected to both entrances of the microfluidic device. Perfluorodecalin was used for its hydrophilic properties with PDMS (static contact angle $\theta_s= 23 \pm 1 ^{\circ}$) and since it does not swell PDMS \citep{lee2003solvent}. This fluorocarbon has a dynamic viscosity $ \mu = 5.1\times 10^{-3} $ Pa.s, surface
tension $ \sigma = 19.3 \times 10^{-3} $ N/m and density $ \rho = 1.9 \times 10^{3} $ kg/m$^{3}$. Then, air is blown in the channel at low pressure to bring the liquid plug to the center of the microfluidic channel and stopped manually when the target position is reached. Finally, the plug motion is forced with either a \textit{unidirectional} or \textit{cyclic} periodic pressure forcing with the MFCS programmable pressure controller. For cyclic forcing one entrance and the exit of the channel are connected to two channels of the MFCS pressure controller. Then an overpressure (compared to atmospheric pressure) is applied alternatively to each end of the channel while the other is set to atmospheric pressure. The resulting shapes of the pressure forcing measured with an internal pressure sensors in these two cases are represented on \cref{Pressure_forcing}. For cyclic forcing the period was fixed to $2T = 4$ s or $2T = 6$ s, with $T$ the duration of a half cycle. The measured unidirectional pressure driving can be approximated by the following analytical expression based on Gompertz functions: 

\begin{equation}
P_t = 900 e^{-3 e^{- 10t}},
\end{equation}

while the cyclic forcing can be approximated by the expression :

\begin{eqnarray}
& & \Delta P_t = 1200e^{-6 e^{-3t}} \mbox{Pa } \mbox{ for } t \in [0;T] \\
& & \Delta P_t = {(-1)}^{n} (P_c - P_d) \mbox{ for } t \in [nT;(n+1)T] \\
& & \mbox{with } P_c = 1200 e^{-2.5e^{-3(t-nT)}} \mbox{Pa } \\
& & \mbox{and }  P_d = 1200 e^{-1.2(t-nT)} e^{-0.02e^{-1.2(t-nT)}} \mbox{Pa }
\end{eqnarray}

Initially the microfluidic channel is dry. Hence, the liquid plug moves on a dry portion of the channel for unidirectional forcing. Nevertheless, the motion of the liquid plug leaves a trailing liquid film on the walls and in the corners of the channel. Thus, for cyclic forcing the liquid plug moves on a prewetted channel after the first cycle as long as it is moves on a portion of the channel already visited by the liquid plug in the previous back and forth motions. 

Experiments are recorded with a Photron SA3 high speed camera mounted on a Z16 Leica Microscope. The resolution of the camera used in the experiments is $1024 \times 64$ pixels, the acquisition frame rate $125$ images/s  and the shutter time $1/3000$ s. The image analysis is then performed using ImageJ software and Matlab. The plug evolution is characterised by monitoring the positions of the rear meniscus $x_r$ and front meniscus $x_f$ (see \cref{setup}), and deducing the evolution of the plug length $L_p(t) = x_f(t) - x_r(t)$ and the speed of the rear meniscus of the plug $U_{r} = dx_{r}/dt$.  

\section{Model of a plug flow in dry and prewetted rectangular microfluidic channels} \label{Model}

The model derived in this paper to describe the dynamics of liquid plugs under pressure forcing combines previous theoretical developments by \cite{baudoin2013airway}, \cite{magniez2016dynamics} and \cite{jfm_baudoin_2018}, and integrates additional elements to include the transition between quasi-static and dynamic liquid film deposition.

\subsection{Dimensional analysis and characterisation of the regime}

In this problem, we consider a single liquid plug of initial length $L_0$ set into motion in a rectangular microfluidic channel under the unidirectional or periodic forcings represented on \cref{Pressure_forcing}. 

\begin{table}[h!]
	\begin{center}
		\begin{tabular}{ccc}
			\hline
			Dimensionless number & Formula & Estimation \\ 
			\hline
			$ \tau _c / \tau_{exp}$ & $l_c/U$ & $  1.8 \times 10^{-2} $   \\
			$ \tau _d / \tau_{exp}$ & $ {{\rho}{{l_c}^2}}/{\mu}$ & $1.3 \times 10^{-2} $  \\
			$Re$ & ${\tau _d} / \tau _c $ & 0.7   \\ 
			$We$ & ${\rho U^2 l_c} / \sigma $ & $1.9 \times 10^{-3}$   \\ 
			$Ca$ & ${\mu U} / \sigma $ & $2.6 \times 10^{-3}$   \\ 
			$Bo$ & ${\rho g h^2} / \sigma $ & $2 \times 10^{-3}$   \\ 
			
			\hline
		\end{tabular}
	\end{center}
	
	\caption{Values of the key dimensionless parameters associated with the mean characteristic velocity $U = 1cm/s$.}
	\label{Table1}
\end{table}

The characteristic parameters in this problem are the width of the microfluidic channel $w$, its height $h$, the viscosity of the liquid plug $\mu$, the surface tension $\sigma$, the speed of the liquid plug $U$ and the characteristic time associated with the plug evolution in the experiments $\tau_{exp}$. For cyclic forcing, this time is simply the half period of the signal $\tau_{exp} = T$, while for unidirectional forcing, it is the time required for the plug to rupture. In the following, the geometry will be characterised by the aspect ratio $\alpha = w/h$ and the characteristic length scale $l_c = \sqrt{wh}$. From these parameters, we can construct the characteristic convection time  $\tau _c = {l_c}/{U} $, and the characteristic viscous diffusion time $ \tau _d = {{\rho}{{l_c}^2}}/{\mu}$. Then we can characterise the flow regime by introducing the following dimensionless numbers: the Reynolds number $(Re = \tau _d / \tau _c) $ which compares convection to viscous diffusion, the Weber number $(We = \rho U^2 l_c / \sigma)$ which compares inertia to surface tension, the capillary number $(Ca = \mu U / \sigma)$ which compares viscous effects to surface tension effects, the Bond number which compare gravity to surface tension ($Bo ={\rho g h^2} / \sigma $), and finally the ratio of the experimental characteristic time $\tau_{exp}$ to the convective and diffusion times $\tau_{exp} / \tau_c$ and $\tau_{exp} / \tau_d$. In the experiments the average velocity is typically $U_{mean} = 1$ cm/s and the maximal velocity $U_{max} = 4.5$ cm/s. The time required for the plug to rupture varies between $0.5$ s and $5$ s for unidirectional forcing and the half period of periodic forcing is $T = 2$ s or $T = 3$ s. These values enable the estimation of the dimensionless numbers introduced previously and summarised in  \cref{Table1} (for these estimations, we take $\tau_{exp} = 1$ s and $U = 1$ cm/s).

These values of the dimensionless numbers indicate that, globally, surface tension effects are dominant over viscous effects, themselves being dominant over inertial effects. This means that away from the walls, the shape of the meniscus is mainly dictated by the minimisation of the interfacial capillary energy. This undeformed part of the meniscus is called the static meniscus. Nevertheless, as demonstrated first by Bretherton \citep{bretherton1961} in cylindrical tubes, viscous effects still play an important role close to the walls due to the incompatibility between the no-slip condition (null velocity at the walls) and the homogeneous motion of an undeformed meniscus. This singularity leads to the existence of large shear stresses close to the walls and, hence, the deposition of a thin trailing film behind the rear meniscus and a change in the apparent contact angle for the front meniscus. Thus viscous effects still play an important role close to the walls despite the low value of the capillary number.  Then, the relative importance of unsteady effects can be estimated from the ratios $ \tau_c / \tau_{exp} $ and $ \tau_d / \tau_{exp}$. Since, these two ratios are small, the unsteady terms can be neglected in Navier-Stokes equations even if the plug evolves over time: the flow is quasi-static. Finally, the small value of the Bond number indicates that gravity effects can be neglected.  Based on this analysis, we expect the liquid plugs dynamics in the present problem to be considered as a quasi-static visco-capillary flow 
governed by steady Stokes equation.

\subsection{Model of the plug dynamics.}


The model describing the pressure drop in the microfluidic channel is obtained by equalizing the driving pressure head $\Delta P_t $ (\cref{Pressure_forcing}) to the sum of the pressure drop resulting from viscous dissipation in the bulk of the liquid plug $\Delta P ^{bulk}_{visc}$, the pressure drops at the front and rear meniscus of the plug $\Delta P ^{men}_{front}$, $\Delta P ^{men}_{rear}$ and the pressure drop $\Delta P_{bubble}$ inside the air.  Simple estimation of these pressure drops show that this latest contribution can be ignored compared to the other ones (see e.g. \cite{kreutzer2005inertial}). Thus, the steady state balance of pressure across the liquid plug becomes: 
\begin{equation}
\Delta P_t = \Delta P ^{bulk}_{visc} + \Delta P ^{men}_{front}  + \Delta P ^{men}_{rear} 
\end{equation}

\subsubsection{Viscous pressure drop}

The pressure drop resulting from a laminar flow of a fluid in a rectangular geometry is given by \citep{mc_white_2003}:
\begin{equation}  \label{bulk}
\Delta P ^{bulk}_{visc} = \dfrac{a \mu Q L}{w h^3}
\end{equation}
with $a = 12 \bigg[   1  - {\dfrac{192}{\pi^5 \alpha}}{\tanh\Bigg(\dfrac{\pi \alpha}{2}\Bigg)}^{-1}  \bigg]$, $L$ the portion of the tube considered and $Q$ the flow rate.  In the limit $\alpha \gg 1$ considered here ($\alpha = 17.5$ in the experiments) and for a liquid plug of length $L_p$, this expression becomes:
\begin{equation}  \label{bulk}
\Delta P ^{bulk}_{visc} = \dfrac{12 \mu Q L_p}{w h^3}
\end{equation}
where $Q = U_r S_r$ is the flow rate, $S_r$ is the the cross sectional area open to air behind the liquid plug and $U_r$ the speed of the rear meniscus. This expression relies on two assumptions: (i) it assumes that the pressure drop inside the plug follows a Hagen-Poiseuille law despite the finite size of the plug and the recirculation occurring close to the menisci and (ii) it assumes the same speed for the front and rear meniscus. The validity of the first approximation has been tested numerically with the OpenFoam Volume of Fluid code \citep{jfm_baudoin_2018} in 2D. These 2D simulations show that equation (\ref{bulk}) is an excellent approximation  of the viscous pressure drop (error $<4.5 \%$) as long as the length of the plug remains larger than the height of the channel. For smaller plugs the discrepancy increases progressively but, in this case, the pressure drops at the menisci strongly dominate over bulk viscous pressure drop leading to minor effects of the error on the overall plug dynamics. The second approximation amounts to neglect the evolution speed of the plug length $d L_p / dt$ compared to the translational speed of rear meniscus $d x_r / dt$, since $dx_f / dt = dx_r / dt + d L_p / dt$. Experimental measurements of the speed of the front and rear meniscus show that this approximation holds within a few percent of accuracy. In the remaining part of the manuscript, we will therefore neglect the difference between the front and rear menisci in the estimation of the pressure drops, and the capillary number $Ca$ will therefore be constructed on the rear meniscus velocity: $Ca = \mu U_r / \sigma$.

\subsubsection{Front meniscus pressure drop}

When a liquid plug is at rest (no pressure head) in a rectangular tube, its front and rear menisci adopt complex shapes minimising the interfacial energy. This minimisation problem can be solved with the method of Lagrange multipliers. The solution is a constant mean curvature (CMC) surface verifying the wetting conditions at the walls.  In a rectangular geometry, this shape is rather complex \citep{wong1995motion1} and there is no analytical expression of these surfaces geometry. Nevertheless, what matters when we consider the motion of a liquid plug is not the static shape of the meniscus but the departure from this static shape when the plug moves. Indeed the Laplace pressure jumps resulting from the curvatures of the front and rear meniscus at rest compensate one another leading to a zero contribution. Therefore we will only consider in the following the dynamic pressure jumps at the meniscus, that is to say the Young-Laplace pressure jumps when the plug is moving minus their value at rest.

The computation of the dynamic pressure jumps at the front meniscus can be greatly simplified in a rectangular geometry with high aspect ratio $\alpha \gg 1$. In this case, the principal curvature in one direction $\kappa_h \approx 2 \cos (\theta ^{a}_{d}) / h$ is strongly dominant over the curvature in the other direction $\kappa_w \approx 2 \cos (\theta ^{a}_{d}) / w$, where $\theta ^{a}_{d}$ is the advancing dynamic apparent contact angle. Based on Young-Laplace equation, the dynamic pressure jump at the front meniscus can thus be estimated from the formula: 
\begin{equation}
\Delta P ^{men}_{front} \approx -{\sigma} (\kappa_h - \kappa_h^s), \mbox{ with } \kappa_h \approx 2 \cos (\theta ^{a}_{d}) / h 
\end{equation}
and $\kappa_h^s \approx 2/ h$  the principal curvature in the vertical direction at rest.
In the limit of low capillary numbers, asymptotic expansion leads to:
$cos {\theta ^{a}_{d}} \sim (1 - {{\theta ^{a}_{d}}^2}/2) $, and the dynamic pressure drop becomes:
\begin{equation}
\label{pfront}
\Delta P ^{men}_{front} = \frac{2 \sigma}{h}  \bigg({ \dfrac{{{\theta ^{a}_{d}}^2}}{2}  }  \bigg)
\end{equation}
Of course, this expression is an approximation since (i) it neglects the horizontal curvature compared to the vertical one and (ii) it neglects the thickness of the prewetting film (if the plug is moving on a prewetted capillary tube). This expression is therefore only valid for high aspects ratios $h/w$, low capillary numbers and thin prewetting films.

The next step is to determine the value of the dynamic apparent contact angle $\theta ^{a}_{d}$ as a function of the capillary number. \textit{On a dry substrate}, the dynamic contact angle can be estimated from Hoffman-Tanner's law:
\begin{equation}
\label{hoffman}
\theta ^{a}_{d} = E {{Ca}^{1/3}} 
\end{equation}
with $E$ a constant of order $(4-5)$ for a cylindrical dry tube as reported by \cite{hoffman1975study} and \cite{tanner1979spreading}. \cite{bico2001falling} measured the value of $E$ for a wetting silicon oil liquid plug and \cite{jfm_baudoin_2018} for a perfluorodecalin liquid plug in cylindrical glass capillary tube and obtained consistent values of this parameter: $E=4.3$ and $E=4.4$ respectively. \cite{ody2007transport} and \cite{baudoin2013airway} measured a value of $E=4.9$ for a perfluorodecalin liquid plug moving in PDMS rectangular capillary tubes with high aspect ratio. This value is in good agreement with the experiments performed in this paper. 

A theoretical expression of the dynamic contact angle \textit{on a prewetted surface} was proposed theoretically by \cite{chebbi2003deformation} and validated  experimentally by \cite{magniez2016dynamics} for the motion of a liquid plug in a \textit{prewetted} cylindrical capillary tube:
\begin{equation} \label{tan_theta}
tan {\theta ^{a}_{d}} = {{(3 Ca)}^{1/3}}  f( {{(3 Ca)}^{-2/3}}  cos{\theta ^{a}_{d}} \; h_f/R)
\end{equation}
with $ h_f/R$ the thickness of the liquid film ahead of the liquid plug, $f(y) = \sum_{j=0}^{3} b_n {[log_{10} y]}^n$ and the coefficients $b_n$  are tabulated in \cite{chebbi2003deformation}. This expression enables to integrate a lubrication effect induced by the presence of a prewetting film, which facilitates the displacement of the front meniscus and thus reduces the pressure jump. To the best of our knowledge, no rigorous derivation of an analytical formula exists for rectangular geometries. Nevertheless, a similar expression as equation (\ref{tan_theta}) is expected in rectangular geometries with high aspect ratios $\alpha \gg 1$. Indeed, Chebbi (similarly to Bretherton) derived the above theoretical expression in the approximation of thin prewetting liquid film $h_f$ compared to the radius of the tube ($h_f/R \ll 1$). In this approximation, the radial curvature of the tube is locally neglected and the problem solved is identical to a 2D planar problem. 

In the rectangular configuration, an estimation of the thickness of the prewetting film in the vertical direction is nevertheless missing. To adapt this formula to rectangular geometries, Baudoin et al. \citep{baudoin2013airway} proposed to estimate the relative thickness of the prewetting film by the formula \textcolor{black}{ $\sqrt{1 - \tilde{S}_f}$} with $\tilde{S}_f = S_f / (h \, w)$ the dimensionless cross sectional area open to air in front of the plug. For rectangular channel with large aspect ratios, it is expected that this expression slightly overestimates lubrication effects since the prewetting film is always thicker on the lateral walls than in the central ones. 

Finally, as demonstrated by \cite{jfm_baudoin_2018}, a good approximation of the implicit formula (\ref{tan_theta}) in the limit of low capillary numbers is: 

\begin{equation} \label{eq:theta}
\theta ^{a}_{d} = F Ca^{-1/3}
\end{equation}
with $F = 3^{1/3}  \bigg( b_0 + {b_1} log_{10}(A)  + {b_2} \left[ log_{10}(A) \right]^2 + {b_3} \left[ log_{10} (A) \right]^3 \bigg)$ and $A = {(3 Ca)}^{-2/3} \sqrt{1 - \tilde{S}_f}$. 
This expression shares some similarities with Hoffman-Tanner's law but this time the coefficient $F$ depends on the capillary number, underlining the lubrication effect.

\subsubsection{Rear meniscus pressure drop}

The dynamical pressure drop at the rear meniscus was calculated theoretically by \cite{bretherton1961} in cylindrical geometries, \cite{wong1995motion1,wong1995motion2} for polygonal channels, and later on numerically by \cite{hazel2002steady} for rectangular microfluidic channels of different aspect ratios at finite capillary numbers. Based on the results of \cite{hazel2002steady}, it is possible to infer the following formula for the dynamic pressure jump at the rear meniscus \citep{baudoin2013airway}:
\begin{equation}
\Delta P ^{men}_{rear} = \frac{2 \sigma}{h} D f(\alpha){Ca^{2/3}}
\label{rear}
\end{equation}
with $f(\alpha) = (0.52 + 0.48/\alpha) $ and $D = 4.1$, a constant obtained by \cite{baudoin2013airway} from least square fit of the numerical data points of Fig.8  of \cite{hazel2002steady} in the range $Ca \in [10^{-3},0.3]$, with Broyden-Fletcher-Goldfarb-Shanno (BFGS) minimization algorithm.

\subsubsection{Total pressure drop}

If we combine equations (\ref{bulk}), (\ref{pfront}), (\ref{hoffman}), (\ref{eq:theta}) and (\ref{rear}),  the total pressure drop across the liquid plug becomes:

\begin{eqnarray} \label{Pressure_tot}
\mbox{Dry tube: } \Delta P_t  =  \dfrac{12 \sigma S_r L_p}{w h^3} Ca + \dfrac{\sigma}{h} \left[ E^2  + 2 D f(\alpha) \right] \; Ca^{2/3} \\ 
\mbox{Wet tube: } \Delta P_t =   \dfrac{12 \sigma S_r  L_p}{w h^3} Ca + \dfrac{\sigma}{h} \left[ F^2  + 2 D f(\alpha) \right] \; Ca^{2/3}  \nonumber
\end{eqnarray}

with $E = 4.9$, $F = 3^{1/3}  \bigg( b_0 + {b_1} log_{10}(A)  + {b_2} \left[ log_{10}(A) \right]^2 + {b_3} \left[ log_{10} (A) \right]^3 \bigg)$ and $A = {(3 Ca)}^{-2/3} \sqrt{1 - \tilde{S}_f}$. 
These equations can be written under dimensionless form by introducing the characteristic length $l_c = \sqrt{wh}$ and the characteristic pressure variation $\sigma / h$:

\begin{eqnarray} \label{Pressure_tot_adim}
\mbox{Dry tube: } \Delta \tilde{P}_t  =  12  \sqrt{\alpha} \tilde{S}_r \tilde{L}_pCa + \left[ E^2  + 2 D f(\alpha) \right] \; Ca^{2/3} \\ 
\mbox{Wet tube: } \Delta \tilde{P}_t =   12   \sqrt{\alpha}  \tilde{S}_r  \tilde{L}_p Ca + \left[ F^2  + 2 D f(\alpha) \right] \; Ca^{2/3} \nonumber
\end{eqnarray}
where the tildes indicate dimensionless functions. To achieve a closed set of equations, two equations are missing: one determining the evolution of the plug length $\tilde{L}_p$ and one determining the fluid deposition process on the walls and consequently $\tilde{S}_r$ and $\tilde{S}_f$.

\subsubsection{Evolution of the plug length}

The first equation is simply obtained from a mass balance between the amount of liquid that the plug collects and loses. Let $S_0 = wh$ be the cross section of channel, and $S_r$ and $S_f$ the sections of the tube open to air behind and in front of the plug respectively. The mass balance becomes:

\begin{equation}
{S_0}dL_p = ({S_0}-{S_f})dx_f - ({S_0}-{S_r})dx_r 
\end{equation}
with $dx_r$ and $dL_p$ the displacement of the rear interface and the variation of the plug length during an infinitesimal time step $dt$ and $dx_f = dL_p + dx_r$. Therefore, the equation giving the evolution of the length of the liquid plug takes the form: 
\begin{equation} \label{dl_dt_dim}
\dfrac{dL_p}{dt} =  \bigg[ \dfrac{S_r}{S_f} -1  \bigg] U = \frac{\sigma}{\mu} \bigg[ \dfrac{S_r}{S_f} -1  \bigg] Ca
\end{equation}
where $S_r$ depends only on the dimensionless speed of the rear meniscus $S_r = S_r(Ca)$ and $S_f$ depends on the history of the liquid deposition on the walls and the position of the front meniscus $S_f = S_f(x_f,t)$.
If we introduce the characteristic length $l_c$ and the viscocapillary time scale $\tau = \mu l_c / \sigma$, we obtain the dimensionless equation:
\begin{equation} \label{dl_dt}
\dfrac{d \tilde{L}_p}{d \tilde{t}} =  \bigg[ \dfrac{\tilde{S}_r(Ca)}{\tilde{S}_f(\tilde{x_r},\tilde{t})} -1  \bigg] Ca 
\end{equation}
When the plug moves on a dry surface, then ${\tilde{S}_f} = 1$. Otherwise the value of $\tilde{S}_f$ is either inferred from the initial condition if the plug moves on a prewetted tube or from a memory of the liquid deposition by the liquid plug when the plug undergoes cyclic motion (see \cite{jfm_baudoin_2018} for cylindrical tubes). In many papers, the wet fraction $m$ is introduced instead of the air fraction $S_r$. The wet fraction is the relative portion of the tube section occupied by the liquid. These two parameters are linked by the formula: $m= 1 - \tilde{S_r}$.

\subsubsection{Quasi-static and dynamic wet fraction} \label{Wet_fraction}

\begin{figure}[h!]
	\centerline{\includegraphics[width=12cm,height=2.5cm]{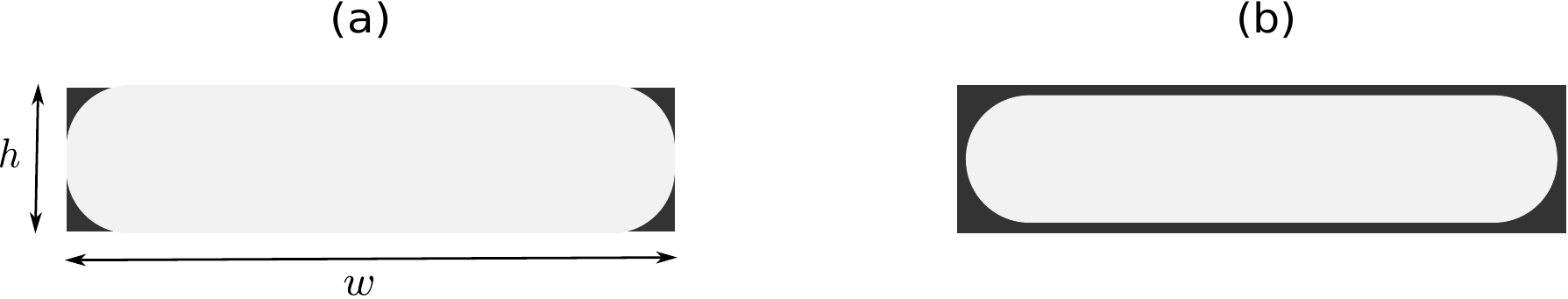}}
	\caption{ Sketch of the cross-section behind the liquid plug (a) when $Ca \rightarrow 0$ (quasi-static film deposition regime) and when (b) $Ca \gtrsim Ca_c$ (dynamic film deposition regime). Black: liquid, white: air.}
	\label{confi_bubble_rec}
\end{figure}

An air finger pushing a liquid plug in a cylindrical tube adopts a cylindrical shape closely fitting the shape of the cylindrical tube. Such close fitting is not possible in rectangular channels due to the presence of sharp corners. In the low Bond number $(Bo\ll1)$ and low capillary number limit $(Ca\ll1)$, the section of the bubble far behind the rear meniscus can adopt two configurations \citep{de2008steady} schematically represented on \cref{confi_bubble_rec} depending on the value of the capillary number: 
\begin{itemize}
\item (i) When $Ca \rightarrow 0$, in the quasi-static limit, the liquid only covers the corners of the tube and the limit between the liquid and the air are four quarter circles. In this case the wet fraction tends to an asymptotic value called the quasi-static wet fraction $m_s$, which depends only on the aspect ratio of the channel $\alpha$: $m_s = m_s(\alpha)$. Indeed, the liquid deposition in the corners of the tube in this regime mostly relies on the static shape of the rear meniscus, and thus the evolution of the wet fraction with the capillary number is weak.
\item (ii) when $Ca \gtrsim Ca_c$, the limit between the liquid and the air becomes two half circle on the side and a liquid film covers the walls in the center of the channel \citep{de2008steady}. In this case, the wet fraction more strongly depends on the dynamics of the plug and hence on the capillary number. We will call it the dynamic wet fraction $m_d = m_d (Ca,\alpha)$.
\end{itemize}

The transition between these two configurations is progressive (see \cite{de2007scaling}) and occurs at a critical capillary number $Ca_c$ whose value was found in our experiments to lie around $Ca_c=(2 \pm0.1) \times 10^{-3}$, a value coherent with the measurements of \cite{de2007scaling} at similar aspect ratio ($\alpha = 15$): $Ca_c \sim10^{-3}$. 

The \textit{quasi-static wet fraction} $m_s$ in the absence of gravity $(Bo = 0)$ was theoretically predicted in \cite{wong1995motion1}:
\begin{equation} \label{Stat}
m_s = (4-\pi) \tilde{r_s}^2 
\end{equation}
with $\tilde{r_s} = r_s / l_c$ the dimensionless radius of curvature of the four quarter circles delimiting the liquid and the air in the corners of the tube and 
\begin{equation} \label{kappa}
\tilde{r_s} =  \dfrac{\sqrt{\alpha}}{{\alpha +1} + {( ({\alpha - 1})^2 + {\pi \alpha} )}^{1/2} }
\end{equation}
For an aspect ratio $\alpha = 17.5$ such as in our experiments, this formula gives the value $m_s = 0.012$. The wet fraction can be estimated experimentally by monitoring the evolution of the plug size as a function of the rear meniscus velocity when $Ca \ll Ca_c$. We found the average estimate of this parameter (over all the experiments performed and described in the next section) to be: $m_s = 0.03$. This value is larger than the theoretical value predicted by \cite{wong1995motion1}. Nevertheless the theoretical value derived by this author was obtained with zero influence of gravity ($Bo = 0$). For finite values of the Bond number, it was shown experimentally by \cite{de2008steady} (see their figure 4) that gravity tends to significantly increase the static wet fraction (these authors found $m_s \approx 0.065$ for $Bo \sim 1$ and $\alpha = 15$). In our case, though small the Bond number (table \ref{Table1}) is not null, which might explain the larger value of the static wet fraction measured experimentally than expected theoretically.  Another tentative explanation would be that, owing to the finite length of the tube, we did not reached the final equilibrium value. For the simulations, we will therefore adopt the constant value: \begin{equation}
m_s = 1 - \tilde{S_r} = 0.03 \mbox{ for } Ca < Ca_c \label{ms}
\end{equation}
under the critical capillary number.

The evolution of the \textit{dynamic wet fraction} in square and rectangular microfluidic channels was investigated experimentally by \cite{kolb1991coating,thulasidas1995bubble,de2007scaling,fries2008segmented,ijmf_han_2009,ijmf_han_2011} and numerically by \cite{hazel2002steady,de2008steady,kreutzer2005multiphase}. In particular, \cite{de2007scaling,de2008steady} found that the measured and simulated evolution of the dynamic wet fraction $m_d$ as a function of the capillary number collapse for all aspect ratio $\alpha$ providing the introduction of an effective capillary number $\hat{Ca} = \left[ 1 + \alpha^2 / \alpha_t^2 \right] Ca$ with $\alpha_t = 6.4$. Thus, the dynamic wet fraction in a rectangular channel with any aspect ratio can be inferred from the behaviour in a square tube. This scaling subsist even for finite Bond numbers as demonstrated both theoretically and numerically by \cite{de2007scaling,de2008steady}. Of course this scaling is only valid for $Ca > Ca_c$ since otherwise, the fluid deposition does not depend on $Ca$ but strongly depends on $\alpha$. Thus, combining (i) the scaling law proposed by \cite{de2008steady} for the effective capillary number, (ii) Aussilous \& Qu\'{e}r\'{e} law \citep{aussillous2000quick} for the evolution of the wet fraction as a function of the capillary number and (iii) the matching condition at the critical capillary number between the quasi-static and dynamic behaviour gives:

\begin{equation}  \label{wet_frc}
m_d = 1 - \tilde{S}_r = \dfrac{m_s + G \left[ (\hat{Ca}/\hat{Ca}_c)^{2/3}-1 \right] }{\left[ 1 + H ((\hat{Ca}/\hat{Ca}_c)^{2/3}-1) \right] }  \mbox{ for } Ca>Ca_c
\end{equation}
where the coefficients $G = 1.05$ and $H = 1.75$ are obtained from best fit with  \cite{de2007scaling} experimental data. \\

Finally, $\tilde{S_r}$ is taken as $\tilde{S_r} = m_s -1$ when the plug moves at a capillary number lower than $Ca_c$ and $\tilde{S_r} = m_d -1$ above. This is of course an approximation since the experimentally observed transition between the dynamic and quasi-static regime at $Ca \sim Ca_c$ is more progressive \citep{de2007scaling}.

\subsection{Numerical resolution of the equations}

The closed set of equations \eqref{Pressure_tot_adim}, \eqref{dl_dt}, \eqref{ms} and \eqref{wet_frc}  are solved using a first order Euler explicit scheme to predict the speed and the evolution of the length of the liquid plug. To cope with the strong acceleration of the liquid plug, an adaptive time step refinement is used: the spatial displacement $\Delta \tilde{x} = x_r^{n+1} - x_r^n$ is kept constant and thus the time step $\Delta \tilde{t}^n$ at iteration $n$ is calculated from the formula: $\Delta \tilde{t}^n = \Delta \tilde{x} Ca^{n-1}$. Convergence on $\Delta \tilde{x}$ has been verified for all the simulations provided in this paper.

\section{Effect of the transition between quasi-static and dynamic film deposition on the dynamics of a liquid plug driven by a unidirectional forcing} \label{Normal_cascade} 

\subsection{Direct experimental evidence of the transition}

\begin{figure}
	\centering
    \includegraphics[width=\linewidth , height=5cm]{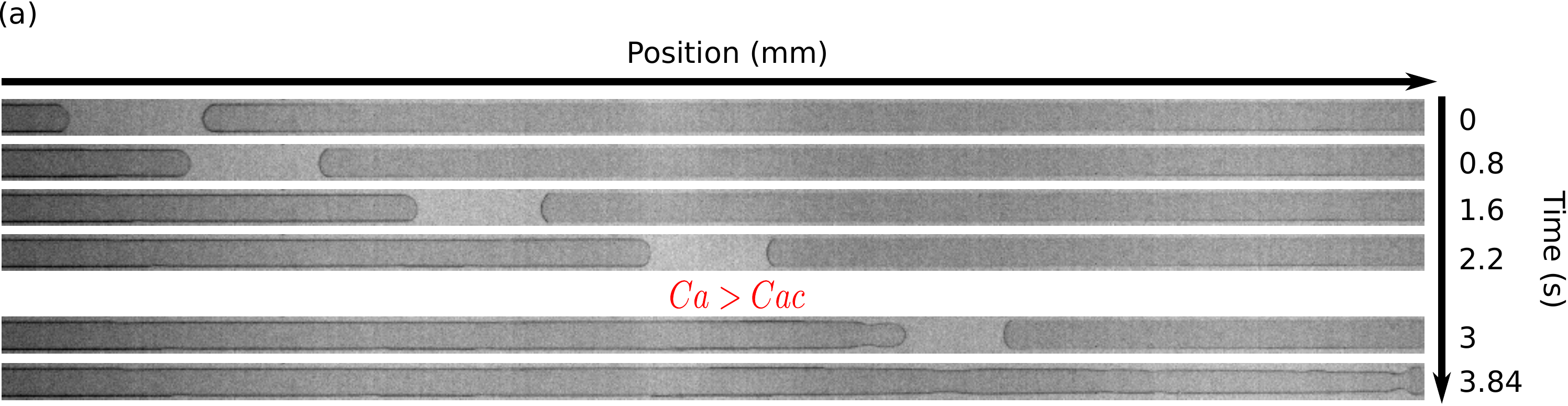}
	\begin{subfigure}[htbp]{0.45\textwidth}
		\includegraphics[width=1.2\linewidth , height=5cm]{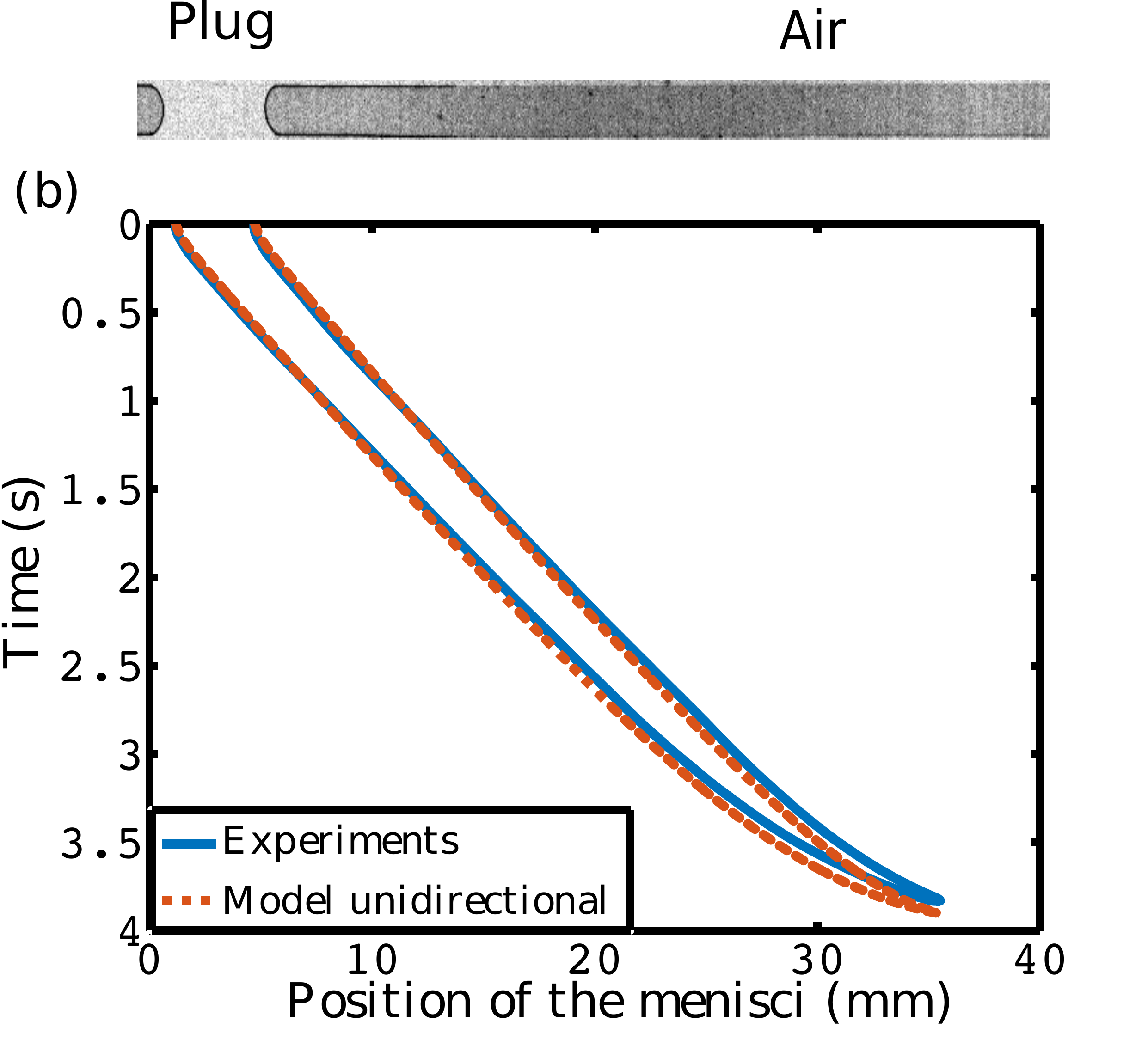}
	\end{subfigure}
	\hfill
	\begin{minipage}[htbp]{0.45\textwidth}
		\begin{subfigure}[b]{\linewidth}
			\includegraphics[width=\linewidth,height=5cm]{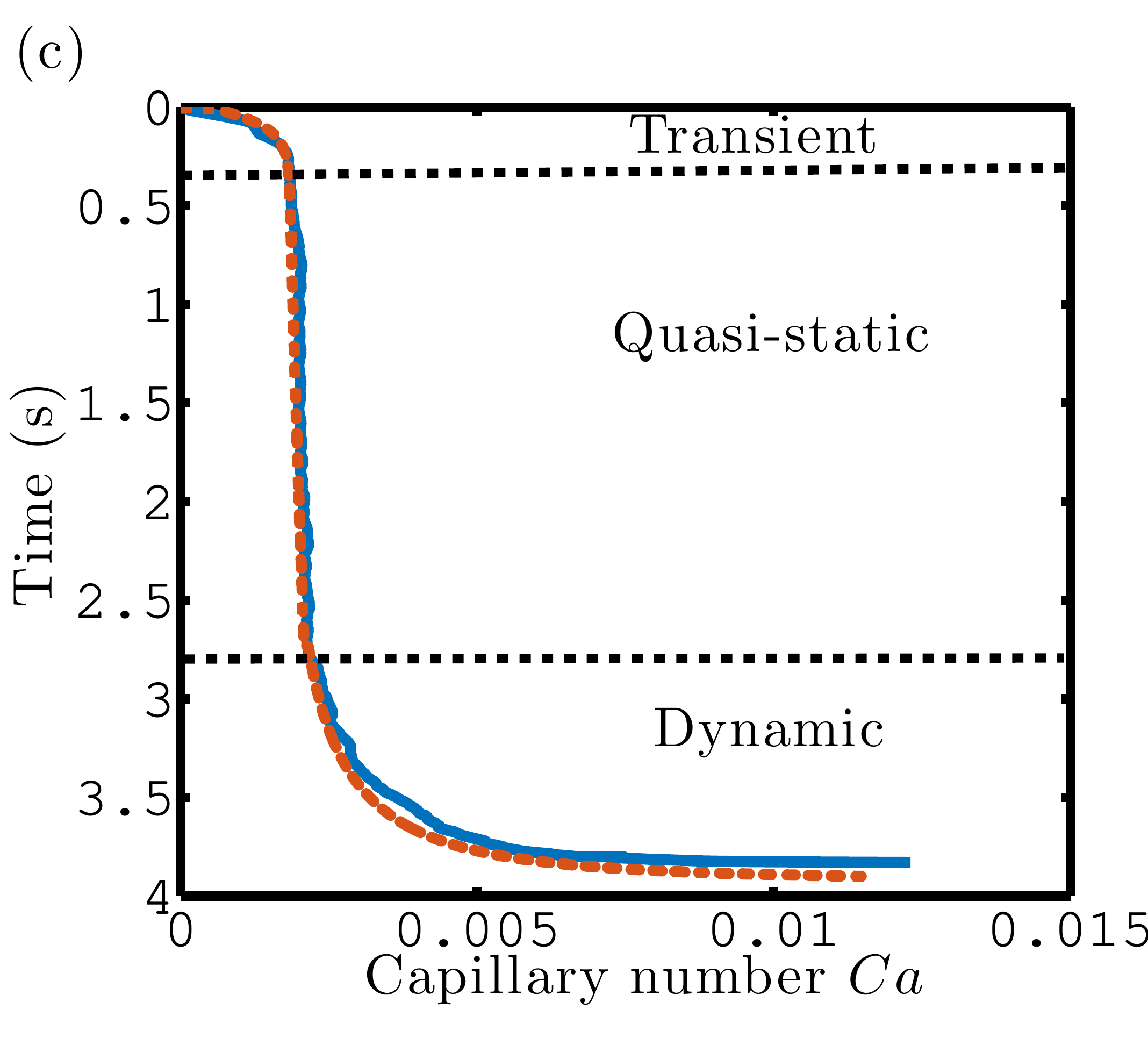}
		\end{subfigure}\\[\baselineskip]
		\begin{subfigure}[b]{\linewidth}
			\includegraphics[width=\linewidth,height=5cm]{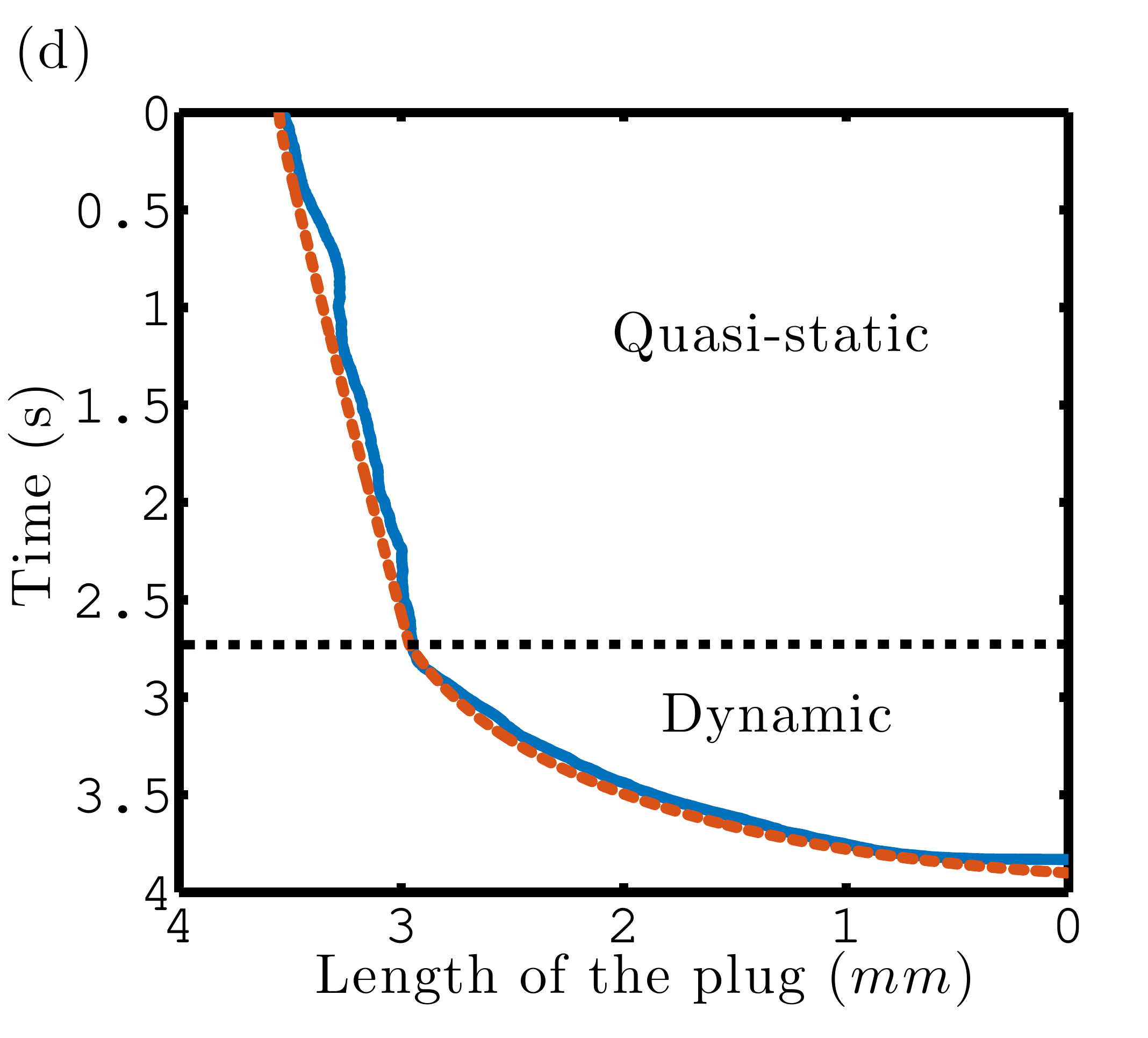}
		\end{subfigure}
	\end{minipage}
\caption{ \label{Normal_cascale2plugs} Dynamics of a liquid plug of initial length $L_0=3.55$ mm pushed in a dry rectangular microfluidic channel with a unidirectional pressure head $\Delta P_t = 1000e^{-3e^{-10t}} $ Pa. (a) Stack showing the evolution of the plug below and above the critical capillary number $Ca_c$. Liquid (air) appears light (dark) grey. (b) Position of the rear and front meniscus as a function of time. (c) Evolution of the capillary number as a function of time. (d) Evolution of the plug length as a function of time. Blue solid curves correspond to experimental measurements and red dashed curves to simulations.}
\end{figure}

The transition between the quasi-static and dynamic liquid film deposition and the associated changes in the plug dynamics are evidenced on \cref{Normal_cascale2plugs} a-d. These figures illustrate the evolution of a liquid plug of initial size $L_o = 3.5$ mm driven by  a unidirectional pressure head $\Delta P_t = 1000e^{-3e^{-10t}}$ Pa. Blue curves correspond to experimental measurements while red curves correspond to simulations with the model developed in the previous section. Figure \ref{Normal_cascale2plugs}a is obtained by stacking snapshots of the plug evolution every $8$ ms when the capillary number lies below its critical value $Ca_c$ and then later on when Ca exceeds $Ca_c$. When $Ca < Ca_c$ no liquid film is visible on the tube lateral sides since liquid  deposition only occurs in the corners of the channel, while this film is clearly visible when $Ca$ exceeds $Ca_c$. Figure \ref{Normal_cascale2plugs}b shows the position of the front and rear menisci as a function of time. Figure \ref{Normal_cascale2plugs}c shows the evolution of the capillary number. The black dashed line (also reported on figure \ref{Normal_cascale2plugs}d) marks the transition (at time $t_c \approx 2.8$ s) between the quasi-static and dynamic film deposition regimes. It corresponds to the time when $Ca$ reaches the critical value $Ca_c= 2 \pm 0.1 \times 10^{-3}$. Before $t_c$ and after the end of the transient regime  ($t>t_t = 0.3$ s) (corresponding to the time required for the pressure controller to achieve a constant value), the increase in the capillary number is very slow. This leads to a quasi-linear variation of the plug size as a function of time as can be seen on figure \ref{Normal_cascale2plugs}d since the wet fraction $m$ is quasi-constant in the quasi-static regime. Then, when the value of the capillary number overcomes the critical value $Ca_c$, the plug undergoes a strong acceleration leading to more and more fluid deposition and eventually to the plug rupture. 

Excellent agreement between the simulations (red) and experiments (blue) is achieved for the evolutions of (i) the position of the menisci (Fig. \ref{Normal_cascale2plugs} b), (ii) the plug dimensionless speed (Fig. \ref{Normal_cascale2plugs} c) and (iii) the plug length (Fig. \ref{Normal_cascale2plugs} d). Our reduced dimension model thus properly captures the main physical ingredients. This model can be used to rationalise the observed tendencies: In the \textit{quasi-static film deposition regime}, the value of the pressure head prescribes an initial value of the capillary number and the size of the plug diminishes quasi-linearly due to film deposition in the corners of the tube. This regular diminution of the plug size leads to a reduction of the viscous resistance of the plug to motion since the viscous pressure drop depends linearly on $L_p$. This induces a slow increase in the liquid plug speed (since the viscous resistance is weak compared to interfacial resistances). Nevertheless, since the wet fraction $m$ does not depend on $Ca$, there is no retroaction of the evolution of the plug speed on the liquid film deposition and thus the evolution remains relatively stable. In the \textit{dynamic film deposition regime} however, the increase in the plug speed leads to more film deposition according to equation \eqref{wet_frc}, itself leading to an acceleration of the plug speed. This retroaction is at the origin of the massive acceleration of the plug and rapid decrease in its size when $Ca$ exceeds $Ca_c$. This behaviour is reminiscent of what is observed in cylindrical tubes \citep{magniez2016dynamics,jfm_baudoin_2018},  while the quasi-static film deposition regime only occurs when there is the presence of sharp corners.

\subsection{Influence of this transition on the plugs rupture time and rupture length}

\begin{figure}[h!]
	\centering
	\begin{subfigure}[b]{0.5\textwidth}
		\includegraphics[width=5.5cm, height=4.5cm]{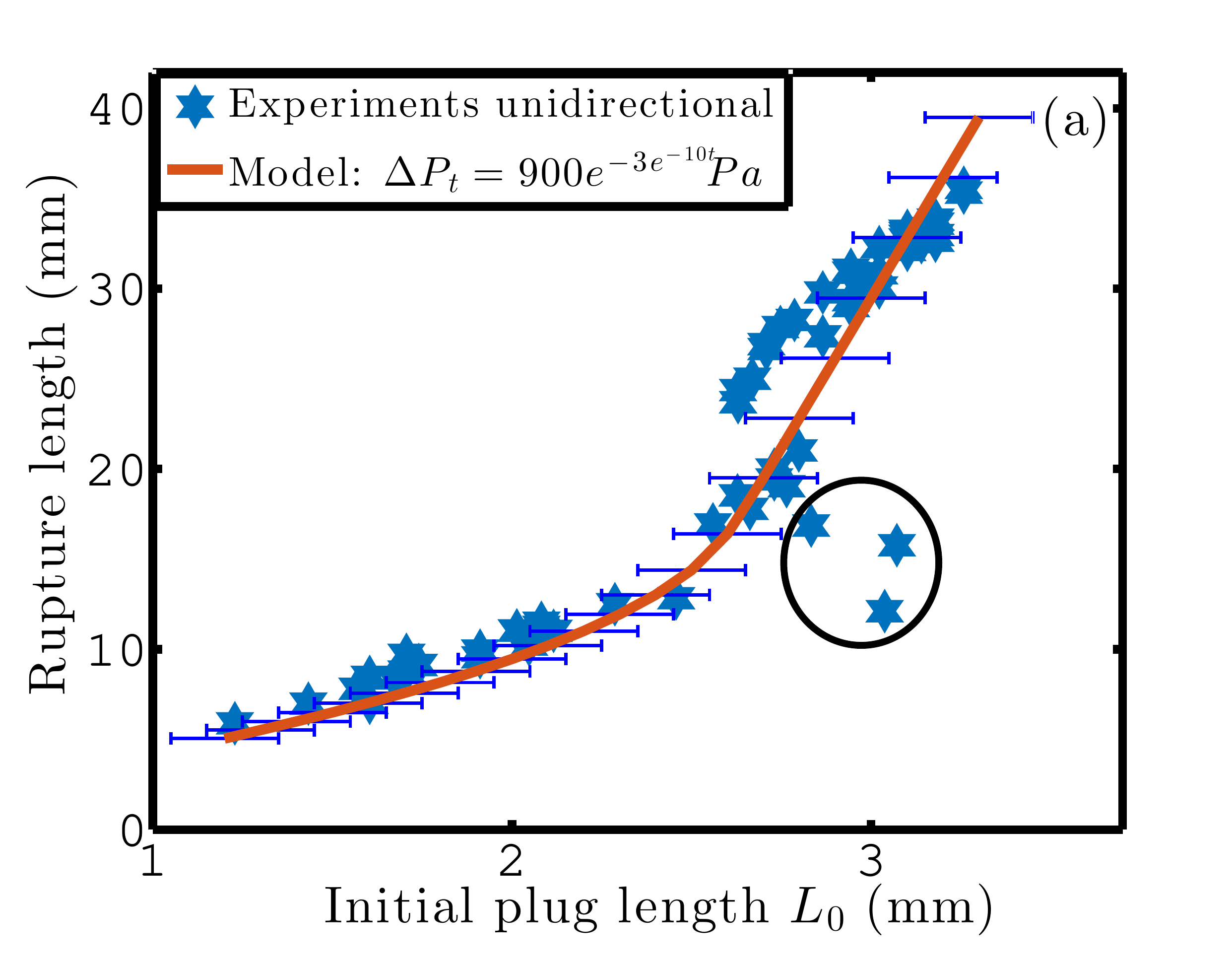}
	\end{subfigure}%
	\begin{subfigure}[b]{0.5\textwidth}
		\includegraphics[width=5.5cm, height=4.5cm]{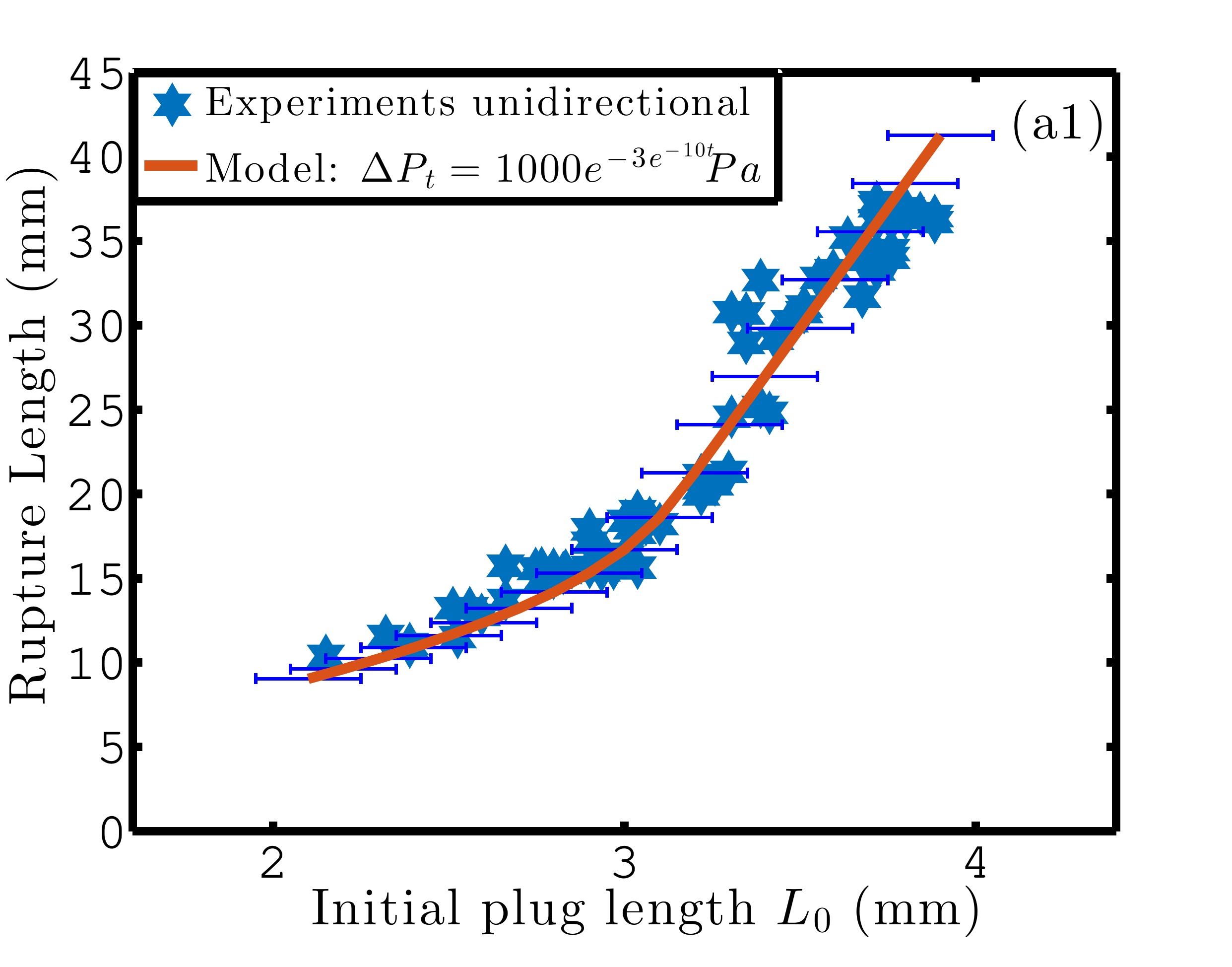}
	\end{subfigure}%
	\\
	\centering
	\begin{subfigure}[b]{0.5\textwidth}
		\includegraphics[width=5.5cm, height=4.5cm]{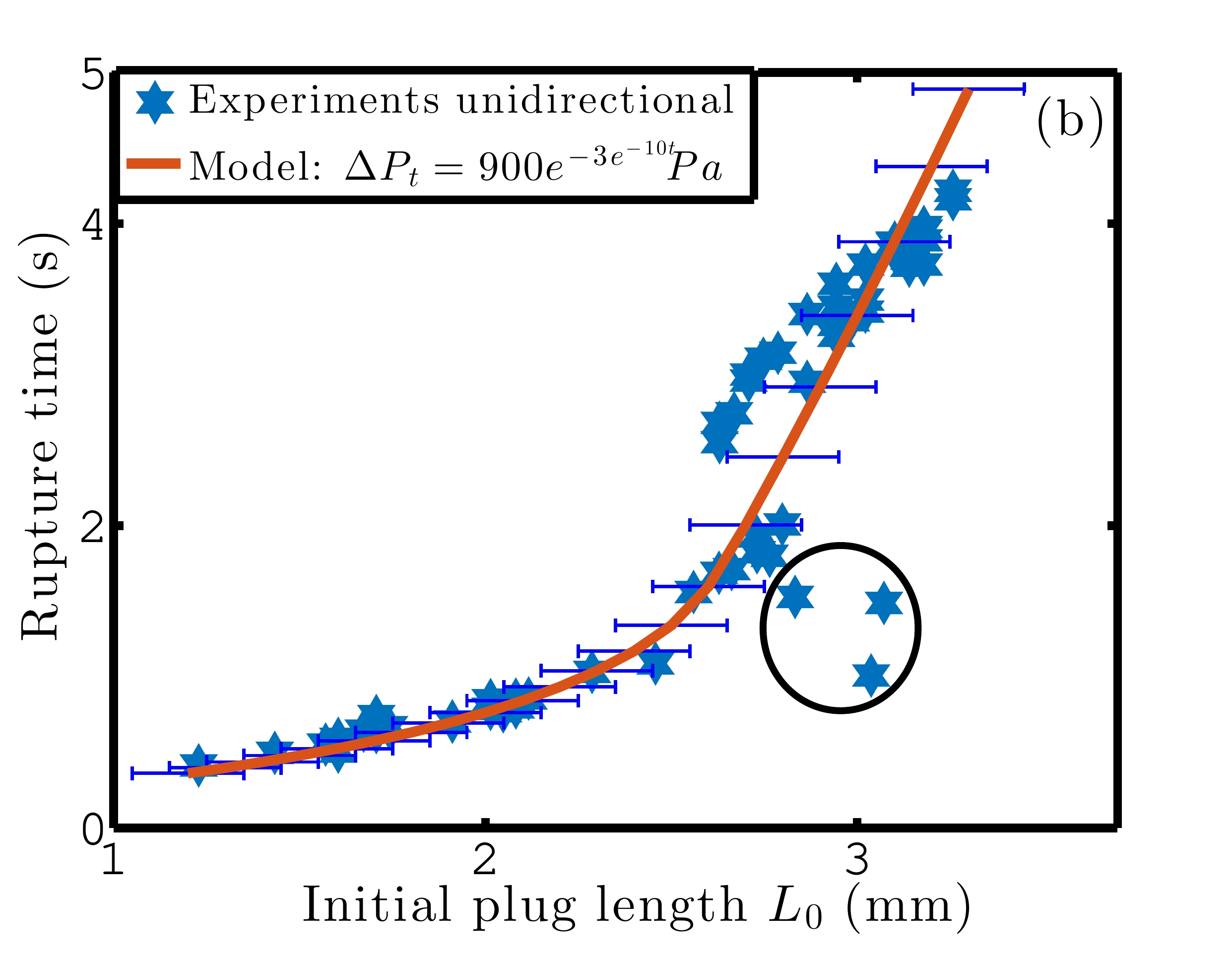}
	\end{subfigure}%
	\begin{subfigure}[b]{0.5\textwidth}
		\includegraphics[width=5.5cm, height=4.5cm]{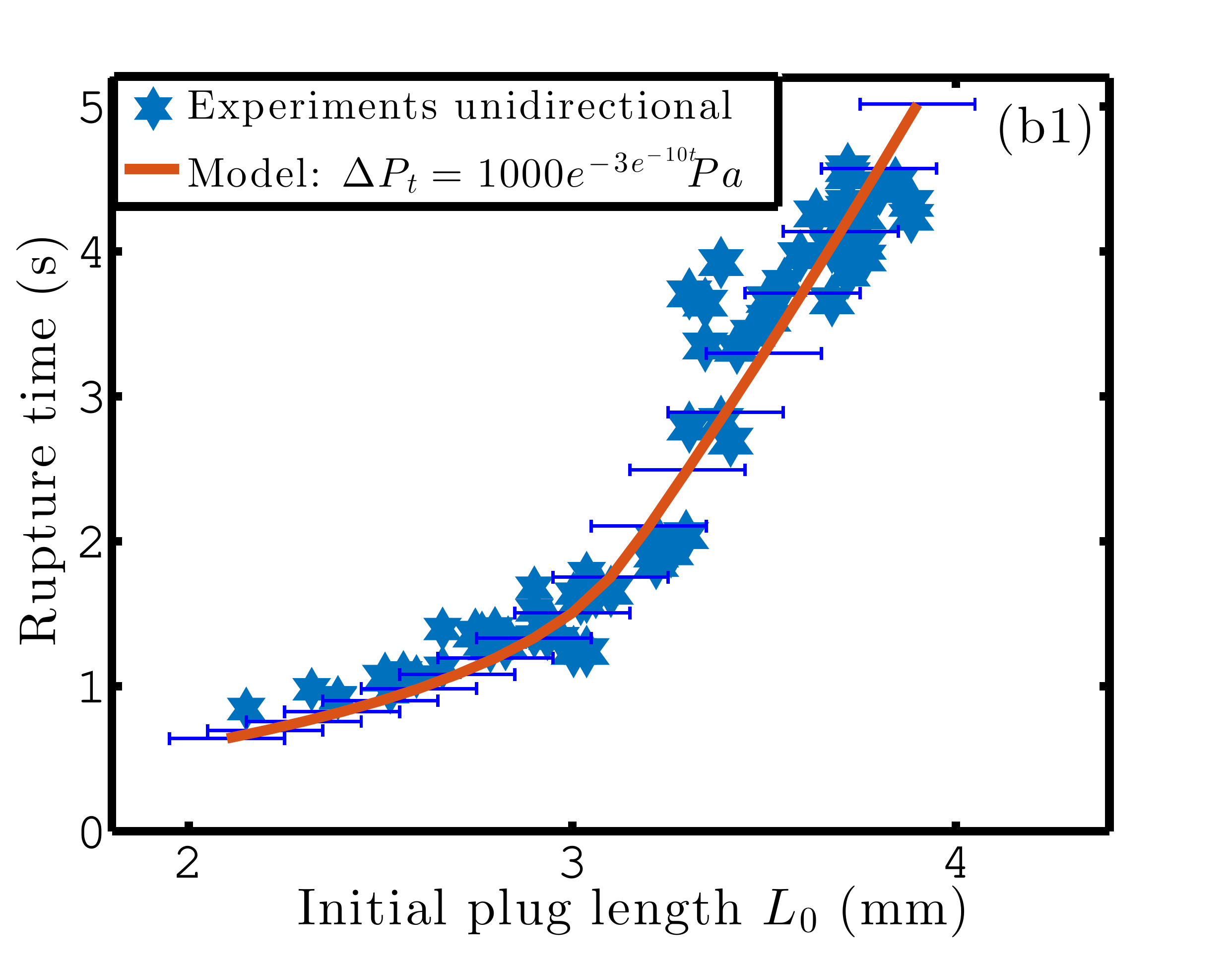}
	\end{subfigure}
	\\
	\centering
	\begin{subfigure}[b]{0.5\textwidth}
		\includegraphics[width=5.5cm, height=4.5cm ]{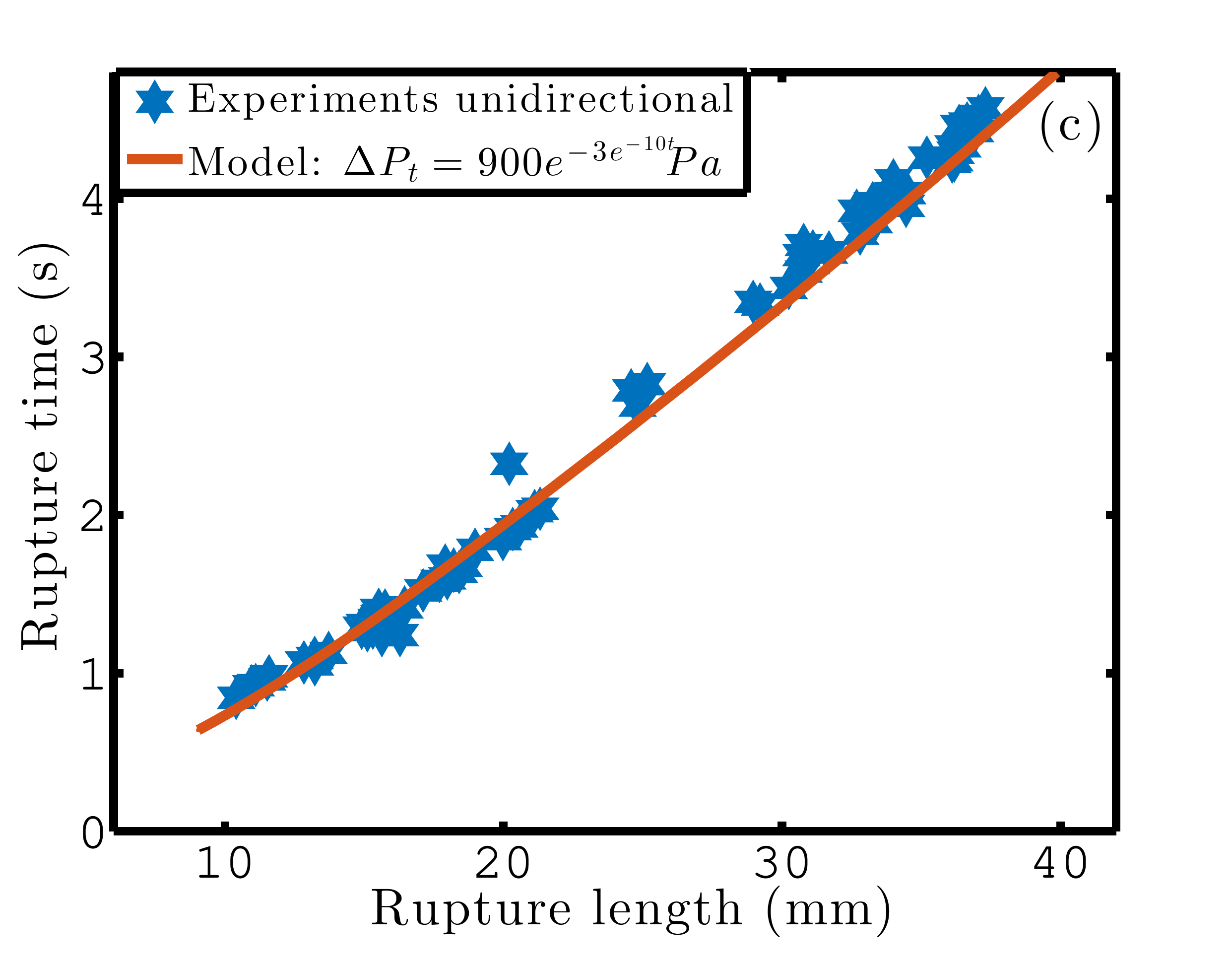}
	\end{subfigure}%
	\begin{subfigure}[b]{0.5\textwidth}
		\includegraphics[width=5.5cm, height=4.5cm]{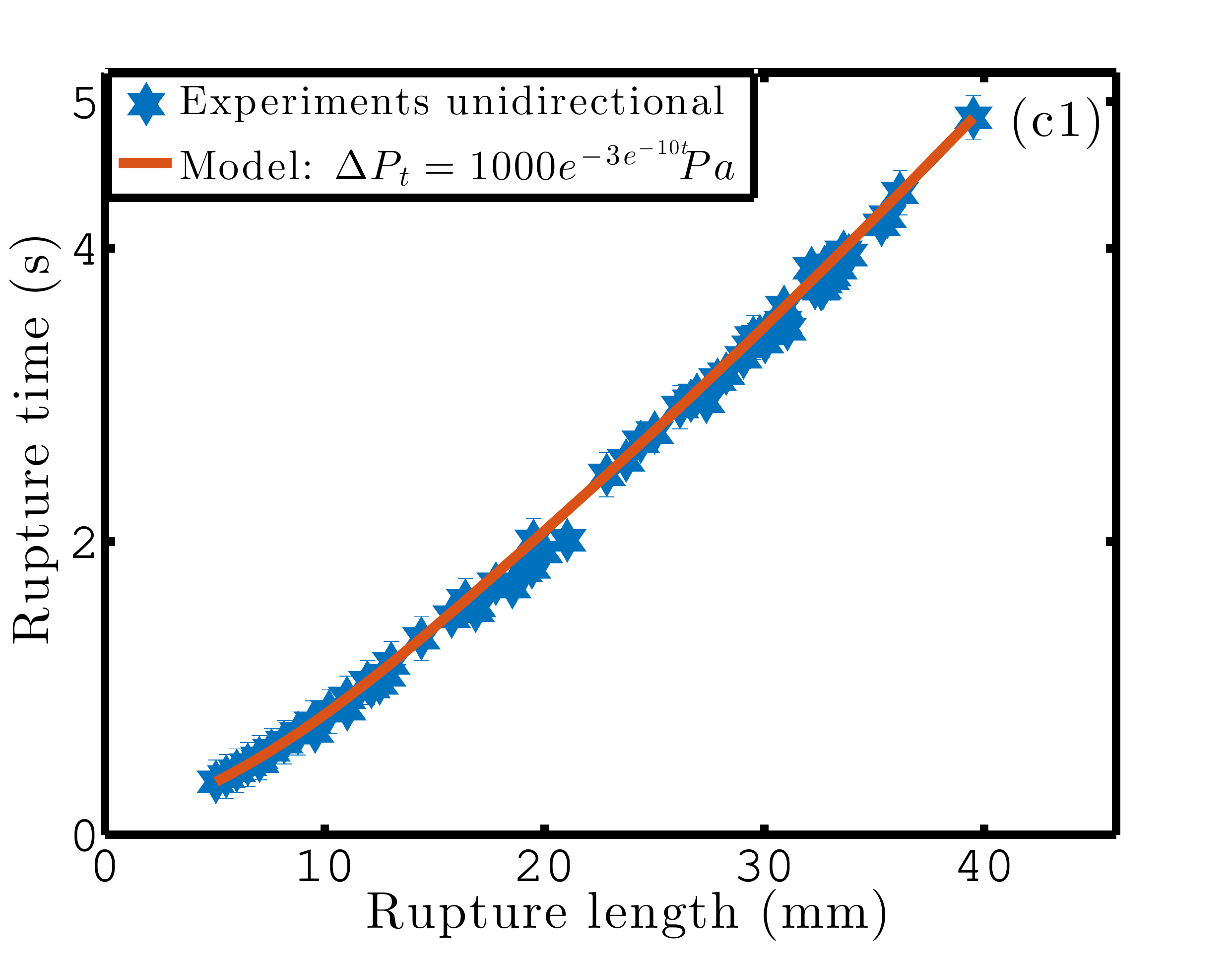}
	\end{subfigure}
	\caption{Rupture lengths (a-a1) and rupture times (b-b1) of a set of liquid plugs as a function their initial lengths when they move on a dry rectangular microfluidic channel under two driving pressure $\Delta P_t = 900e^{-3e^{-10t}}$ Pa (a-b-c) and  $\Delta P_t = 1000e^{-3e^{-10t}}$ Pa (a1-b1-c1). The rupture lengths as a function of the rupture times are given in figures (c-c1). The blue stars correspond to experiments and the red lines to simulations with the model developed in section \ref{Model}. Error bars quantify the error in the determination of the plug initial length owing to the large field of view and the limited resolution of the camera ($1024 \times 64$ pixels). A circle surrounds three data which are out of the global tendency. The most likely reason of this dispersion is that the microfluidic channel was not dried properly and there were some liquid remaining at the front meniscus that lubricated the channel and thus accelerated the plug motion leading to reduced rupture time and length.
}  
   \label{NormalCascade_setofdatas}
\end{figure}

We performed numerous experiments (represented on \cref{NormalCascade_setofdatas}) for different initial plug lengths $L_o$ and two different driving pressures ($\Delta P_t = 900e^{-3e^{-10t}} $ Pa and $\Delta P_t = 1000e^{-3e^{-10t}}$ Pa) to analyse the evolution of the plug rupture time and rupture length in rectangular microchannels. The rupture time is the time elapsed between the start of the pressure head and the plug rupture $(L_p=0)$, that is to say when the front and rear menisci come into contact at the centerline of the channel and the plug breaks, leaving the air flow freely in the channel. The rupture length is the distance traveled by the liquid plug $(D_l = max(x_f)-min(x_r))$ before its rupture. These two parameters quantify the stability of a liquid plug to breaking. The quantitative agreement between experiments (blue stars) and simulations (red lines) enables to validate our model on an an extensive set of experimental data.

Again a transition between two distinct regimes is clearly evidenced on \cref{NormalCascade_setofdatas} (for both the rupture time and the rupture length) at a driving pressure-dependent critical initial plug length $L_o^c$ ($L_o^c \approx 2.6$ mm for  $\Delta P_t = 900 $ Pa and $L_o^c \approx 3.1$ mm for $\Delta P_t = 1000 $ Pa).  Under this critical value of the initial plug length $L_o < L_o^c$, the initial dimensionless plug speed lies above the critical capillary number and thus the dynamics of the liquid plug is only in the dynamic film deposition regime. Thus the plug accelerates rapidly leading to rapid rupture of the plug on a short propagation length scale. Above, this critical initial length $L_o > L_o^c$, the initial capillary number lies under the critical number $Ca_c$ and thus the plug dynamics is initially in the quasi-static film deposition regime. This regime leads to larger plug rupture time and thus propagation distance. Moreover, since in this regime the acceleration is weak, the rupture time and rupture length remain relatively linear function of the plug initial length (see \cref{NormalCascade_setofdatas}). From this analysis, we can infer a theoretical evaluation of the critical initial length $L_o^c$, which delimits the transition between these two regimes. Indeed, $L_o^c$ corresponds to the plug initial length when the initial capillary number is equal to the critical capillary number $Ca_c$. From equation \eqref{Pressure_tot} and by approximating $S_r$ by $wh$ at first order, we obtain:
$$
L_o^c = \frac{h}{12 Ca_c} \left[ \frac{\Delta P_t \, h}{\sigma} - (E^2 + 2 D f(\alpha))Ca_c^{2/3} \right]
$$
This formula gives $L_o^c = 3.1 \pm 0.2$ mm and $L_o^c = 3.5 \pm 0.2$ mm for $\Delta P_t = 900$ Pa and $\Delta P_t = 1000$ Pa respectively. It overestimates by $19 \%$ and $16 \%$ respectively the critical length but nevertheless remains in good agreement with the experimentally measured values. This formula is also consistent with the increase in $L_o^c$ as a function of $\Delta P_t$ observed experimentally. This theoretical prediction of the critical initial length $L_o^c$ is of the upmost practical interest since it enables to predict in which regime will mainly evolve a liquid plug depending on its initial length. An interesting point is also that despite the regime change, the rupture length and rupture time remain relatively proportional to each other (\cref{NormalCascade_setofdatas} c and c1).   

\subsection{Comparison with the dynamics in cylindrical tubes}

\begin{figure}[h!]
	\centering
	\begin{subfigure}[b]{0.5\textwidth}
		\includegraphics[width=6cm, height=5cm]{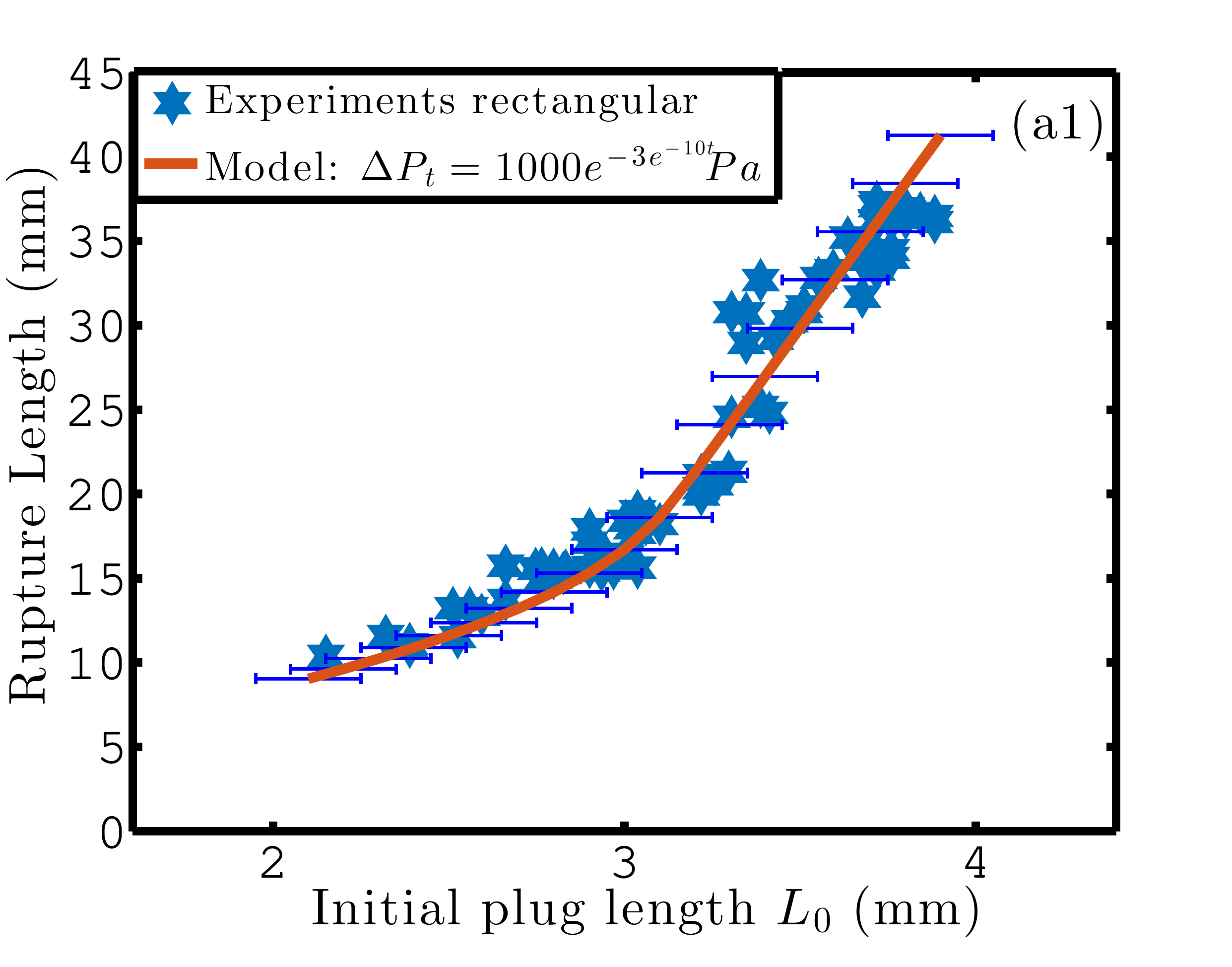}
	\end{subfigure}%
	\begin{subfigure}[b]{0.5\textwidth}
		\includegraphics[width=6cm, height=5cm]{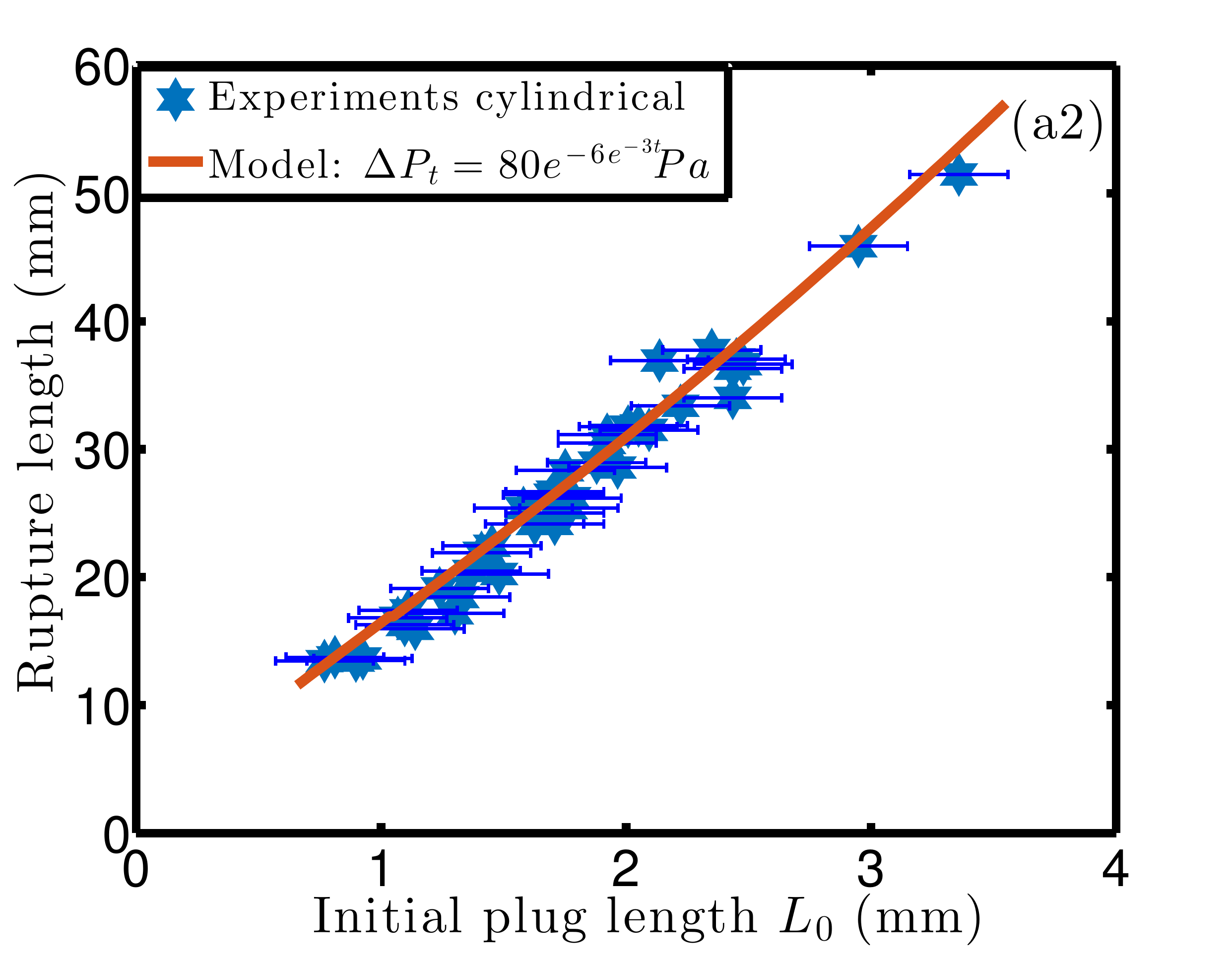}
	\end{subfigure}%
	\\
	\centering
	\begin{subfigure}[b]{0.5\textwidth} 
		\hspace*{0.3mm}
		\includegraphics[width=5.9cm, height=5cm]{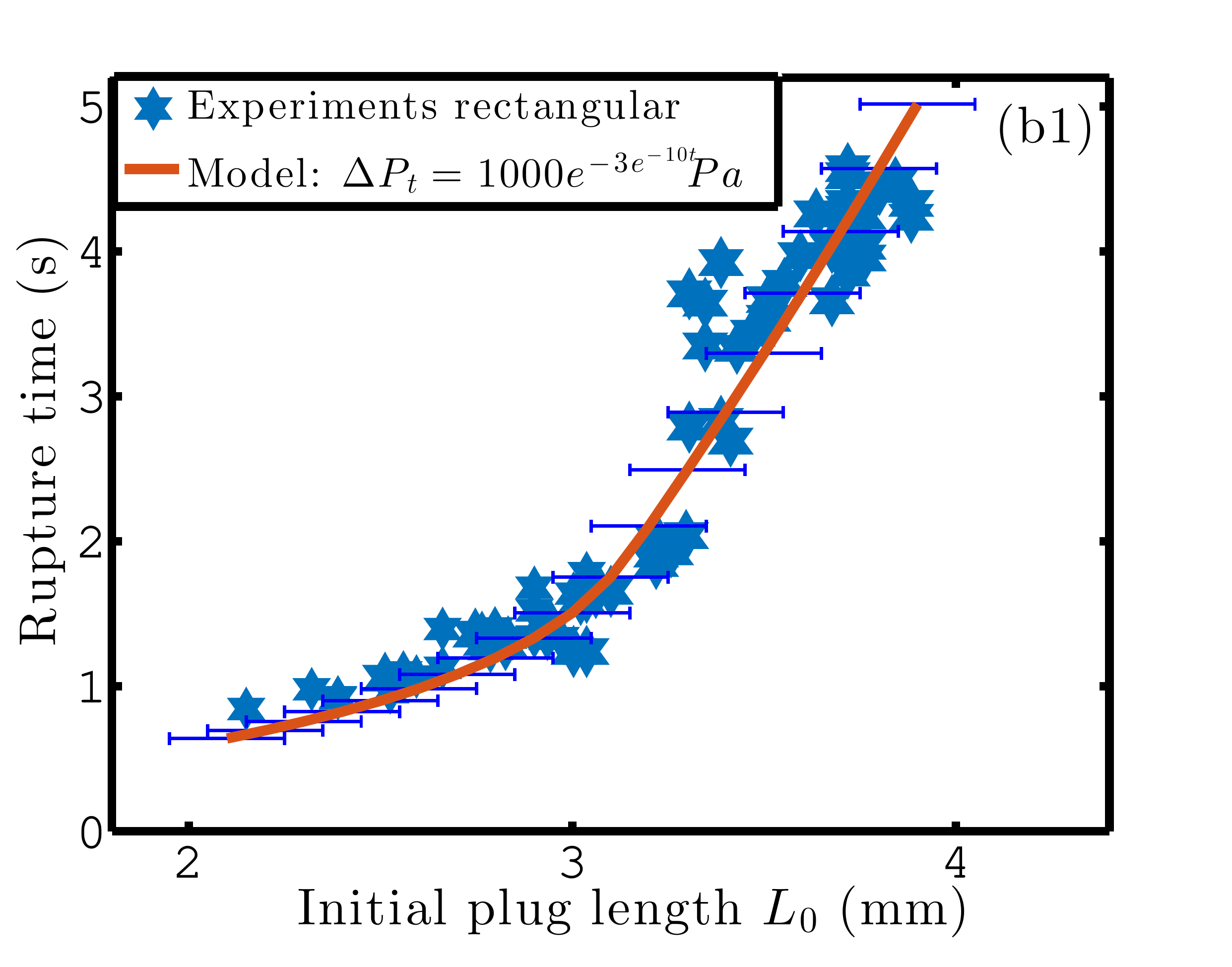}
	\end{subfigure}%
	\begin{subfigure}[b]{0.5\textwidth}
		\includegraphics[width=6cm, height=5cm]{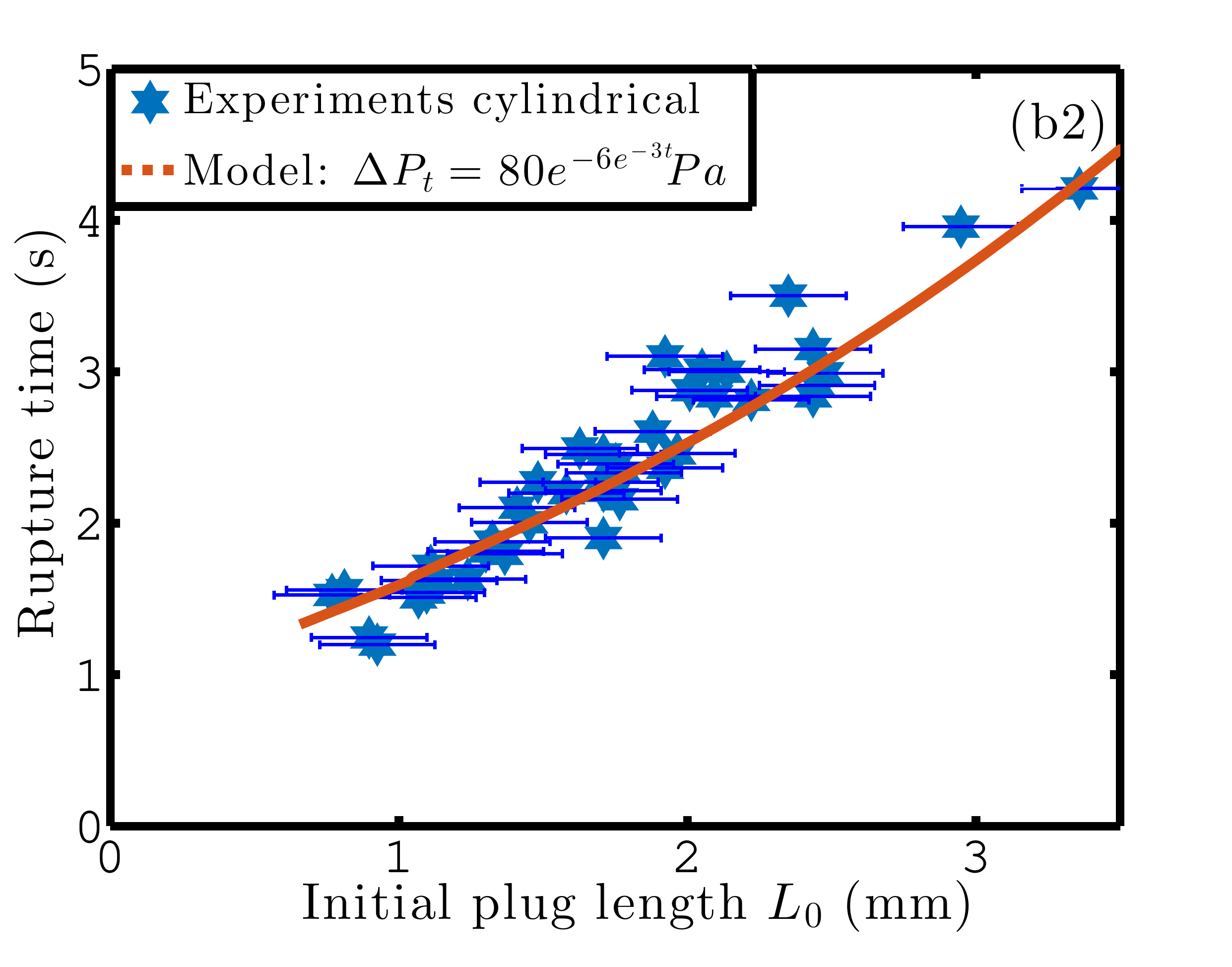}
	\end{subfigure}
	
	\caption{ \label{Setofdatas}  Rupture lengths (a1) and rupture times (b1) of a set of liquid plugs moving in a rectangular microfluidic channel under the pressure driving $\Delta P_t = 1000e^{-3e^{-10t}}$ Pa compared to the rupture lengths (a2) and rupture times (b2) of a set of liquid plugs moving in a cylindrical capillary tube of diameter $D=470 \mu$m under the pressure driving $\Delta P_t = 80e^{-6e^{-3t}}$ Pa (a2-b2) . The blue stars correspond to  experiments and red curves to simulations. Error bars quantify the error in the determination of the plug initial length owing to the large field of view and the limited resolution of the camera ($1024 \times 64$ pixels).
	}  
\end{figure}

Figure \ref{Setofdatas} compares the rupture times and lengths obtained in rectangular and cylindrical tubes in the same range of capillary numbers ($5.5 \times 10^{-5} \lesssim Ca \lesssim 1.2 \times 10^{-2}$). This comparison is not meant to be quantitative since the dimensions of the channels (rectangular: $h = 45 \mu$m, $w = 785 \mu$m, cylindrical: diameter $D = 470 \mu$m) and the driving pressure magnitude ($1000$ Pa for rectangular channels $80$ Pa for cylindrical channels ) are different. The purpose of this figure is only to show that in the same range of capillary numbers, no transition is observed in the case of a cylindrical channel, while a transition is clearly evidenced in a rectangular channel.

\section{Response of liquid plugs to periodic pressure forcings in rectangular microfluidic channels}  \label{Periodic_forcing}

\subsection{Detailed analysis of single plug ruptures.}
\label{Periodic_forcing_3plugs}

\begin{figure}[h!]
	\centering
	\begin{subfigure}[b]{0.33\textwidth}
		\includegraphics[width=\linewidth, height=5cm]{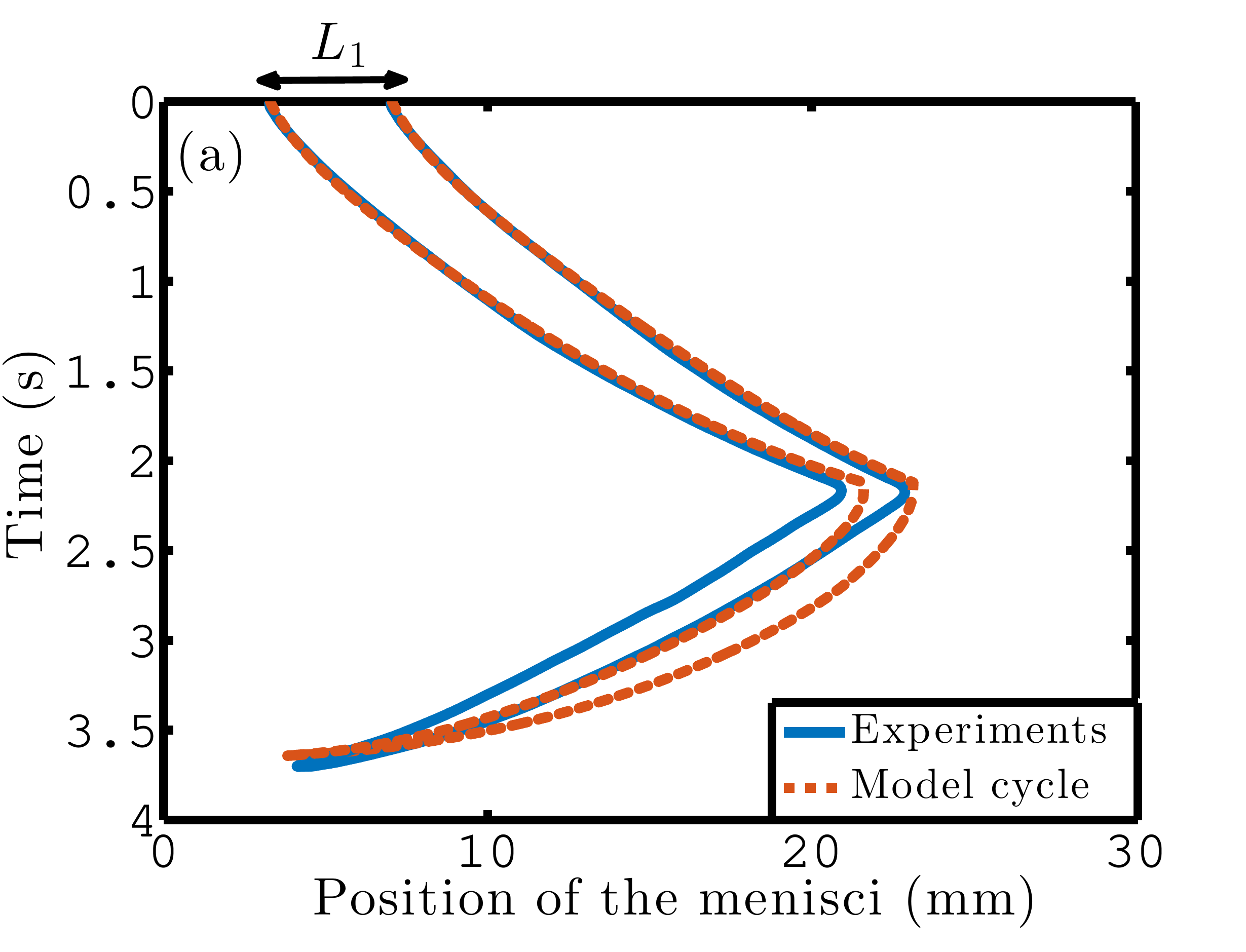}
	\end{subfigure}%
	\begin{subfigure}[b]{0.33\textwidth}
		\includegraphics[width=\linewidth, height=5cm]{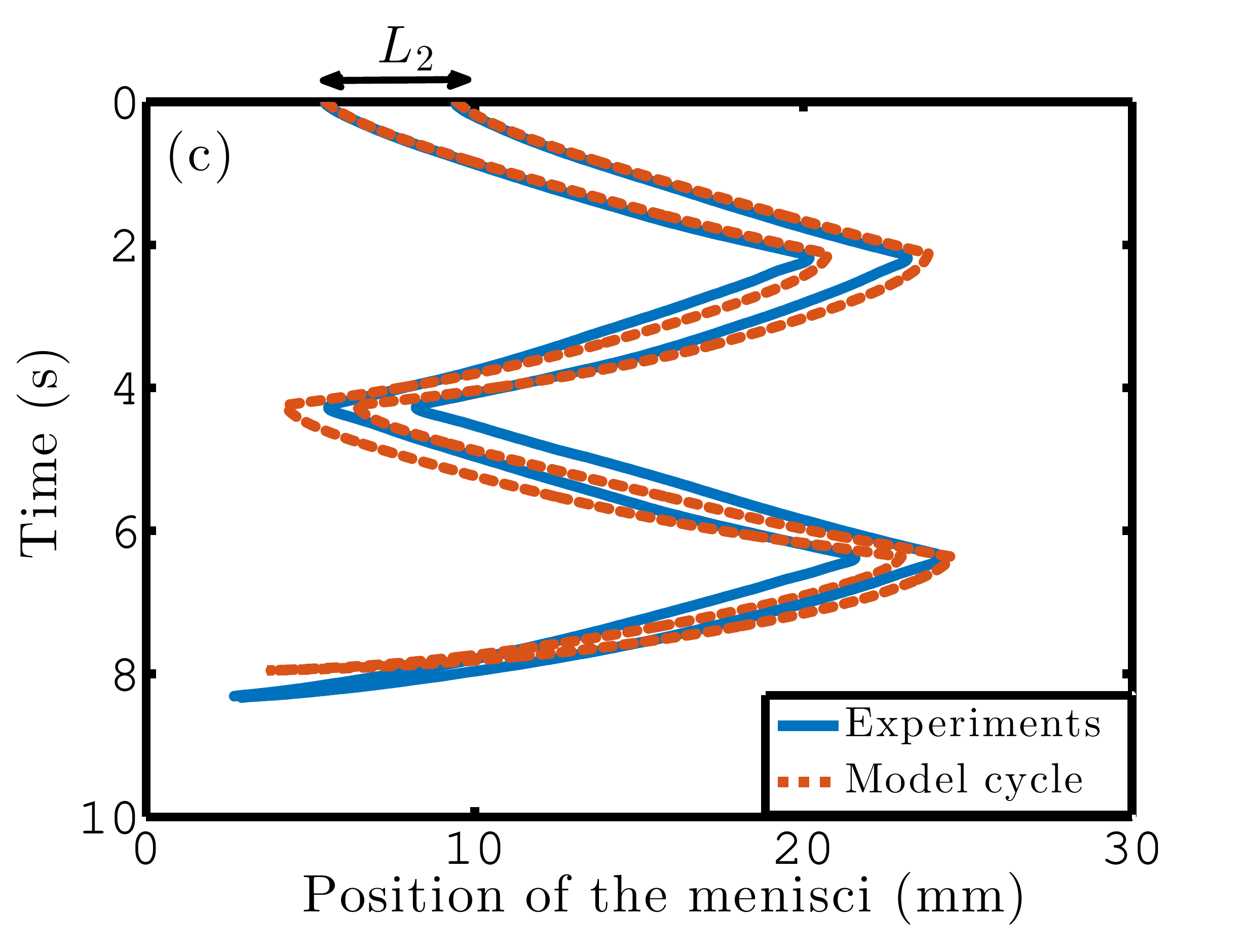}
	\end{subfigure}%
	\begin{subfigure}[b]{0.33\textwidth}
		\includegraphics[width=\linewidth, height=5cm]{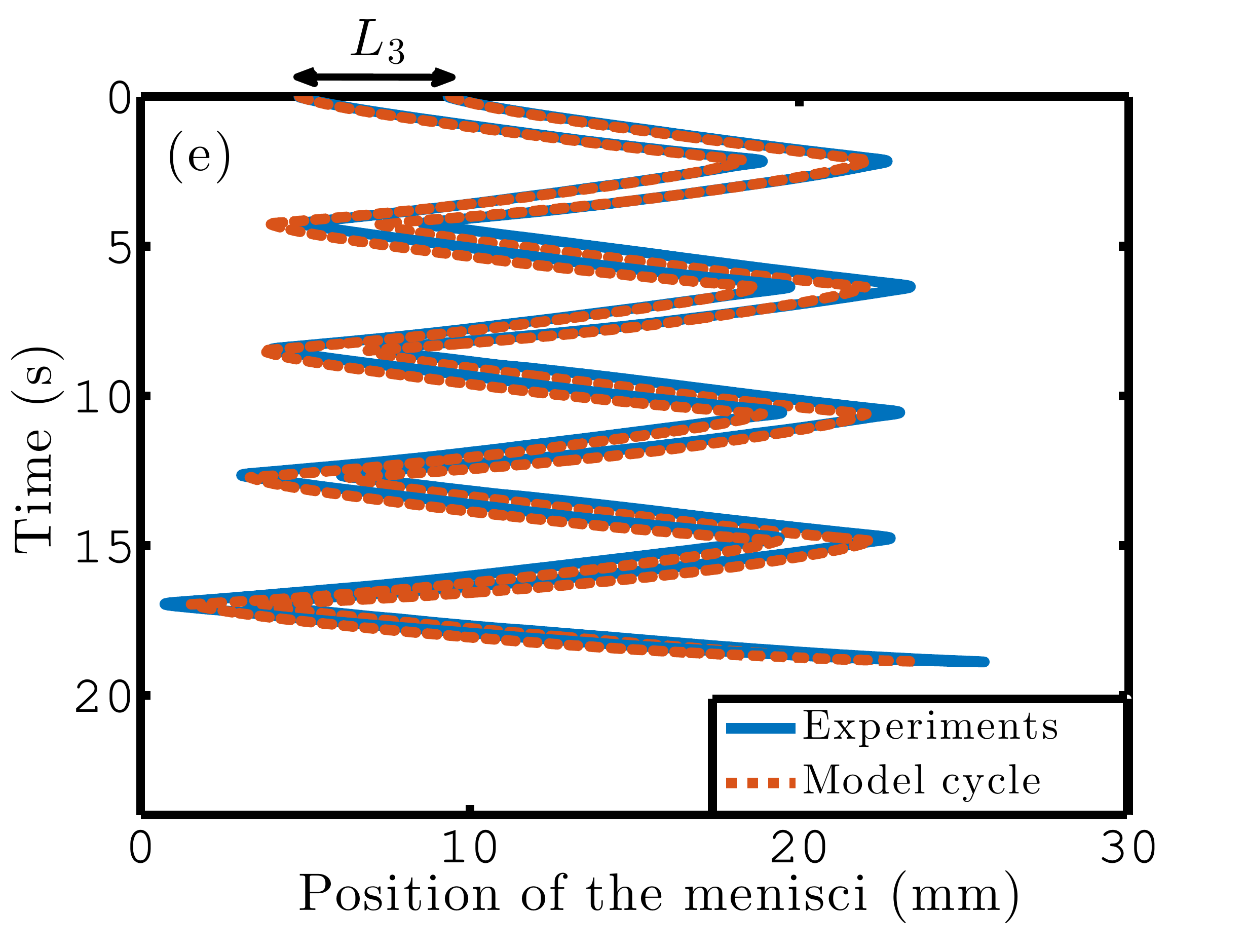}
	\end{subfigure}%
	\\
	\centering
	\begin{subfigure}[b]{0.33\textwidth}
		\includegraphics[width=4cm, height=5cm]{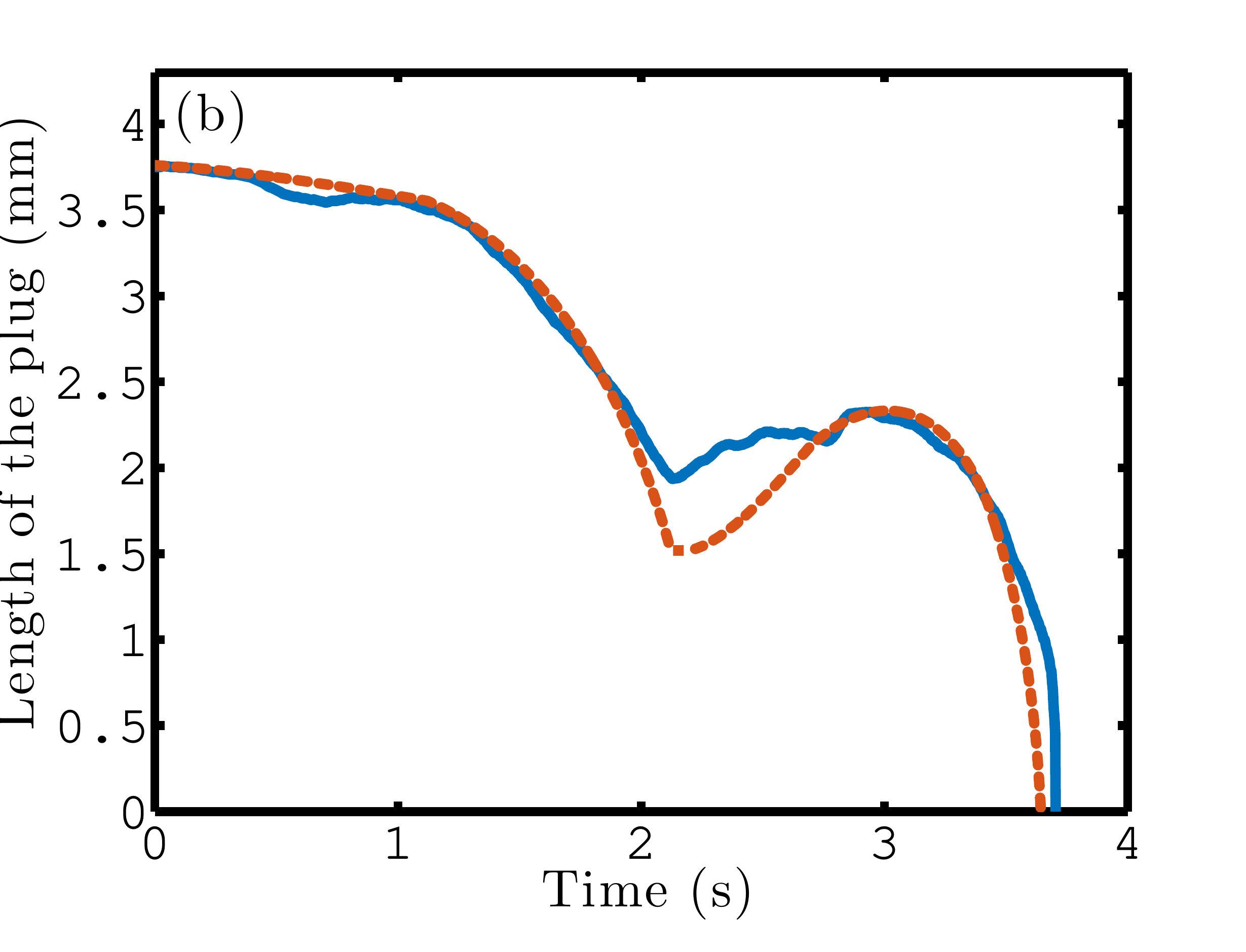}
	\end{subfigure}%
	\begin{subfigure}[b]{0.33\textwidth}
		\includegraphics[width=4cm, height=5cm]{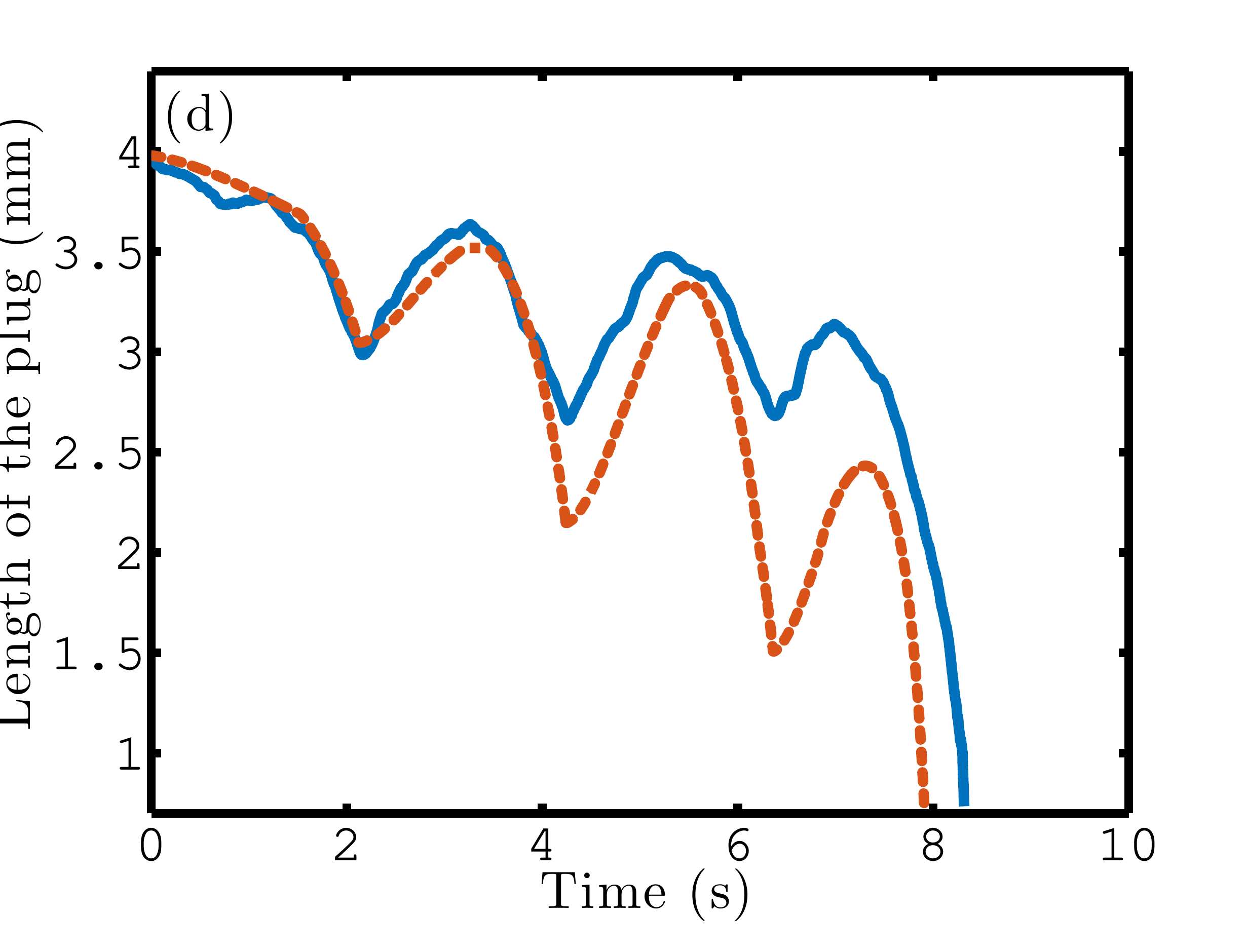}
	\end{subfigure}%
	\begin{subfigure}[b]{0.33\textwidth}
		\includegraphics[width=4cm, height=5cm]{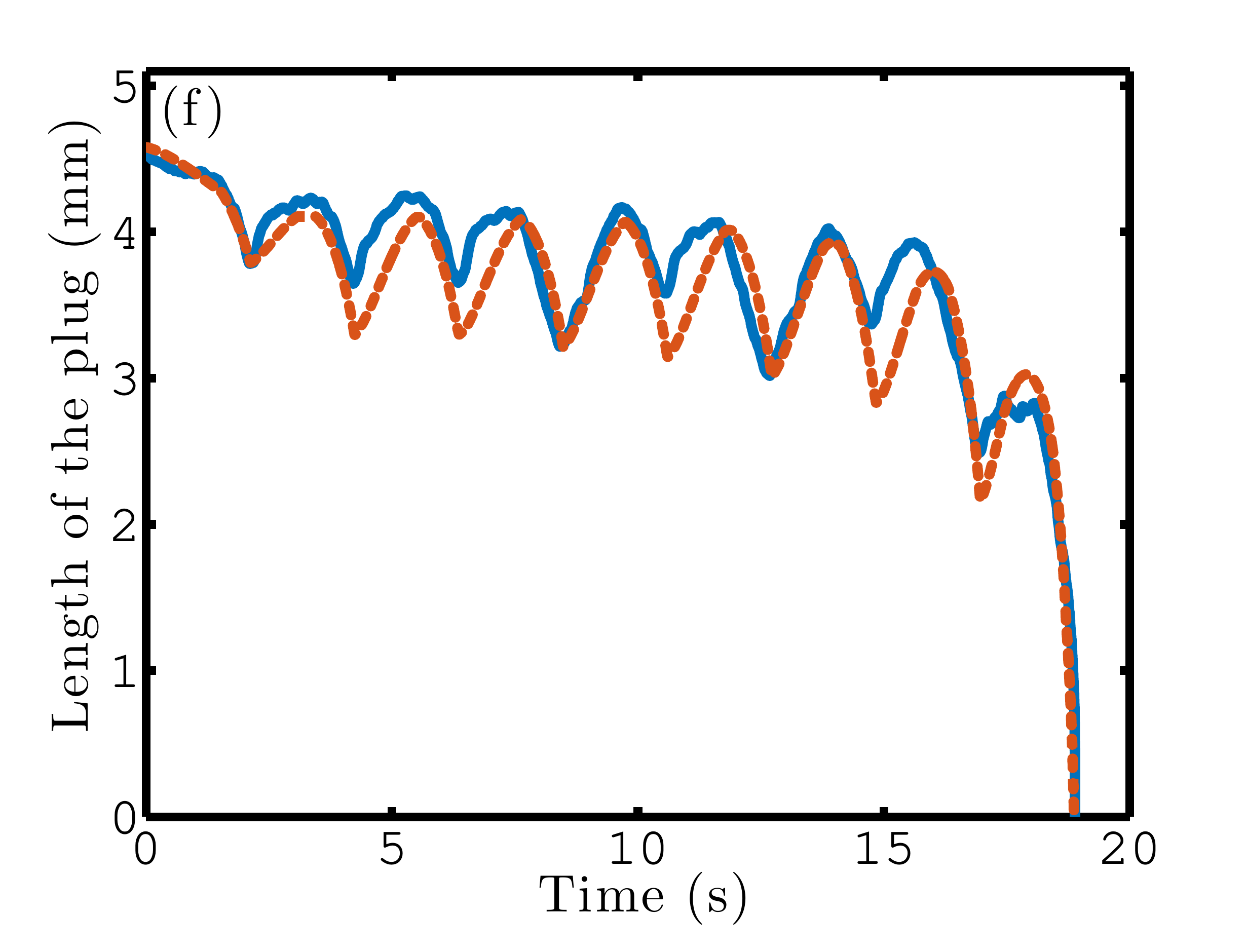}
	\end{subfigure}%
	
	\caption{ \label{Spatiotemporal_cyclic_motion}  Spatiotemporal evolution of three liquid plugs of initial lengths (a-b) $L_1 = 3.85$ mm, (c-d) $L_2 = 4.1$ mm and (e-f) $L_3 = 4.5$ mm driven by the  cyclic pressure forcing represented on \cref{Pressure_forcing}(b). (a-c-e) Positions of the left and right menisci as a function of time. (b-d-f) Evolution of the length of the plug as a function of time.}
 
\end{figure}

We further investigated the response of liquid plugs to cyclic forcing. For this purpose, liquid plugs are inserted at the center of a rectangular microfluidic channel and a cyclic pressure forcing (represented on \cref{Pressure_forcing}) is applied. \cref{Spatiotemporal_cyclic_motion} illustrates the positions of the rear and front menisci  (a-c-e) and  the evolution of the plug length (b-d-f) for three different initial plug lengths: $L_1 = 3.8$ mm (a,b), $L_2 = 4.1$ mm (c,d) and $L_3 = 4.5$ mm (e,f). The blue curves correspond to experiments and the red curves to simulations. For these three initial lengths, the plugs undergo oscillations eventually leading to their rupture. The experimental results show that the evolution of the plug length is not monotonic: the plug size first increases and then decreases during each back and forth motion. This is a consequence of the progressive increase in the driving pressure (\cref{Pressure_forcing}b): at the beginning the driving pressure is low, the plug moves slowly and leaves less liquid behind it than it recovers from the liquid film lying in front of it. Then, when the driving pressure reaches a critical pressure (derived in \cite{magniez2016dynamics} in the case of cylindrical tubes), the tendency is inverted.

The number of oscillations before the plug rupture increases with the initial length of the plug.  For the longest plug ($L_3 = 4.5$ mm), a clear transition can be seen between a first phase where the plug undergoes relatively stable oscillations with weak net evolution of its length from one cycle to another (see \cref{Spatiotemporal_cyclic_motion}f before time $t =15$ s) and a second phase with a brutal acceleration of the plug rapidly leading to its rupture ($t\geq 15$ s)

\subsection{Specificity of the cyclic dynamics of liquid plugs in rectangular channels compared to cylindrical channels.}

\begin{figure} 
	\centering
	\begin{subfigure}[b]{0.5\textwidth}
		\includegraphics[width=6cm, height=5cm]{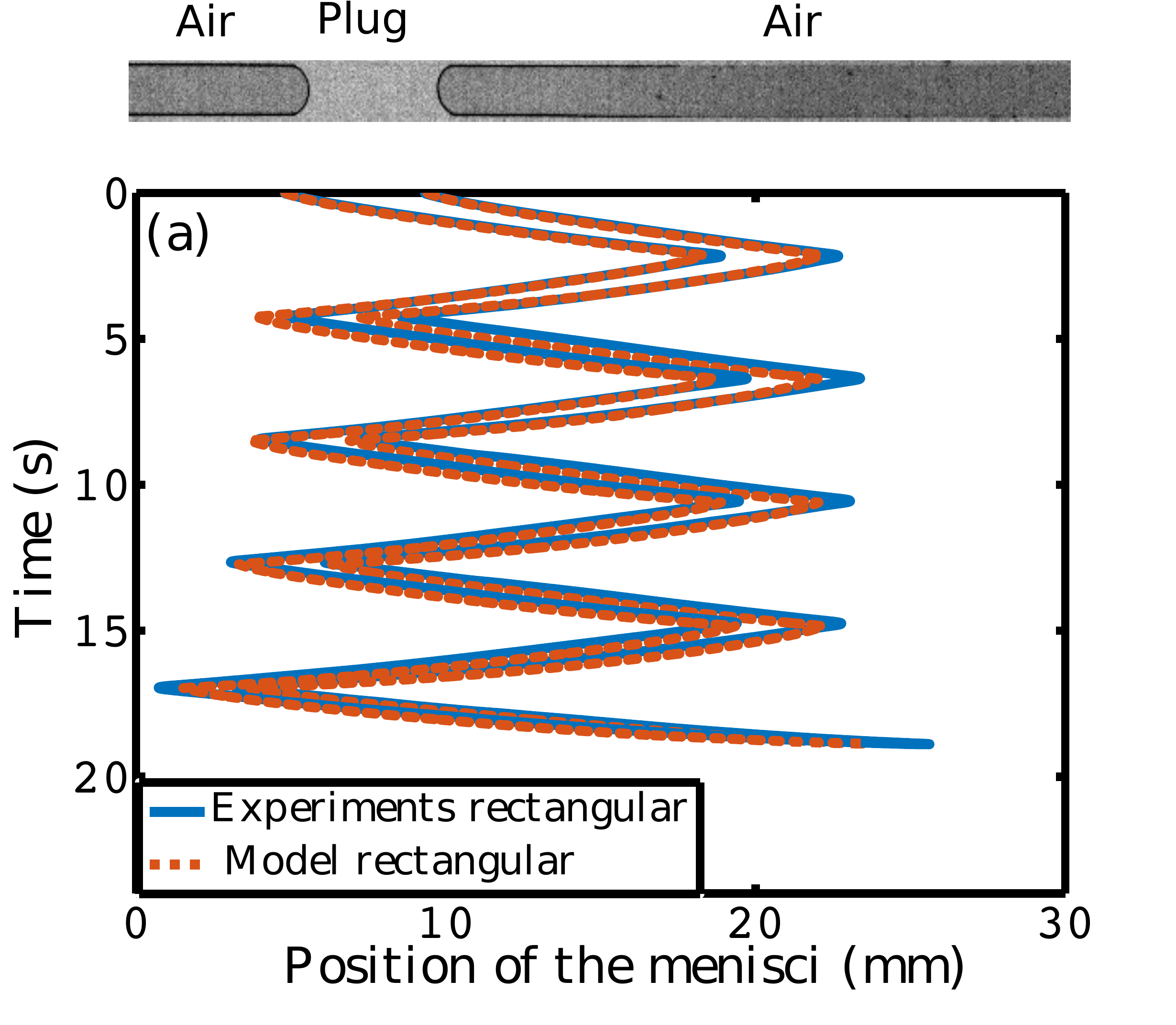}
	\end{subfigure}%
	\begin{subfigure}[b]{0.5\textwidth}
		\includegraphics[width=6cm, height=5cm]{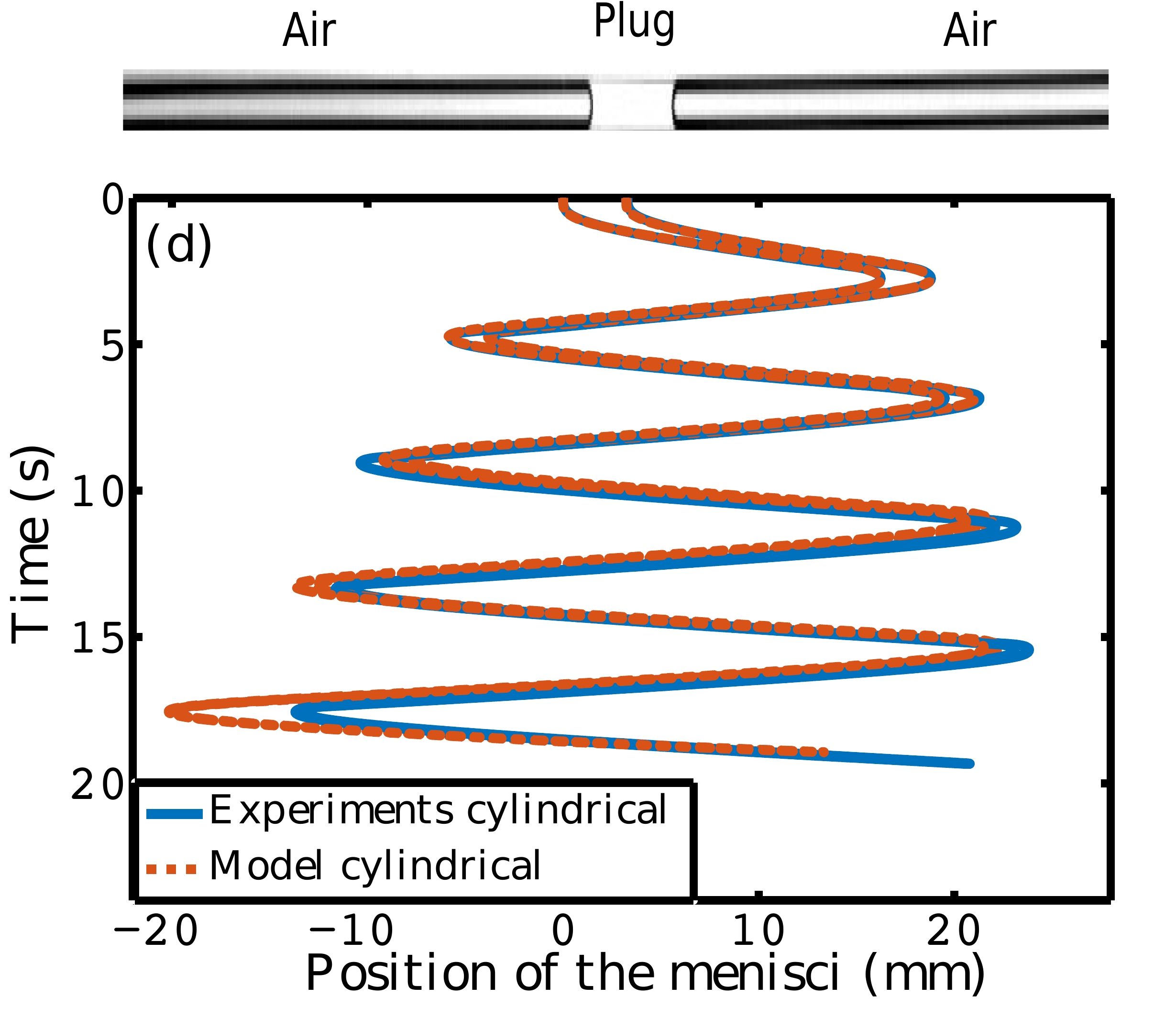}
	\end{subfigure}%
	\\
	\centering
	\begin{subfigure}[b]{0.5\textwidth}
		\includegraphics[width=6cm, height=5cm]{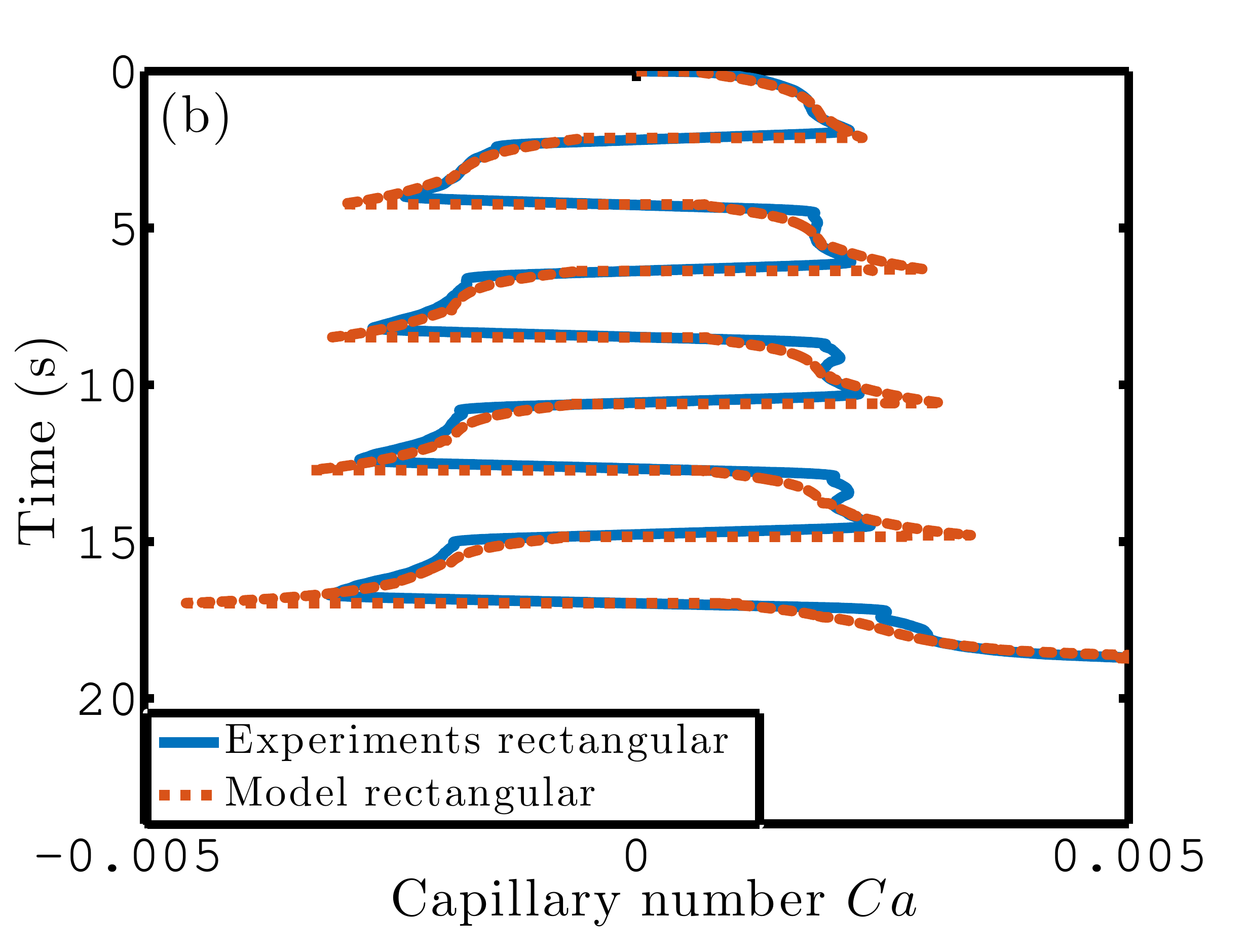}
	\end{subfigure}%
	\begin{subfigure}[b]{0.5\textwidth}
		\includegraphics[width=6cm, height=5cm]{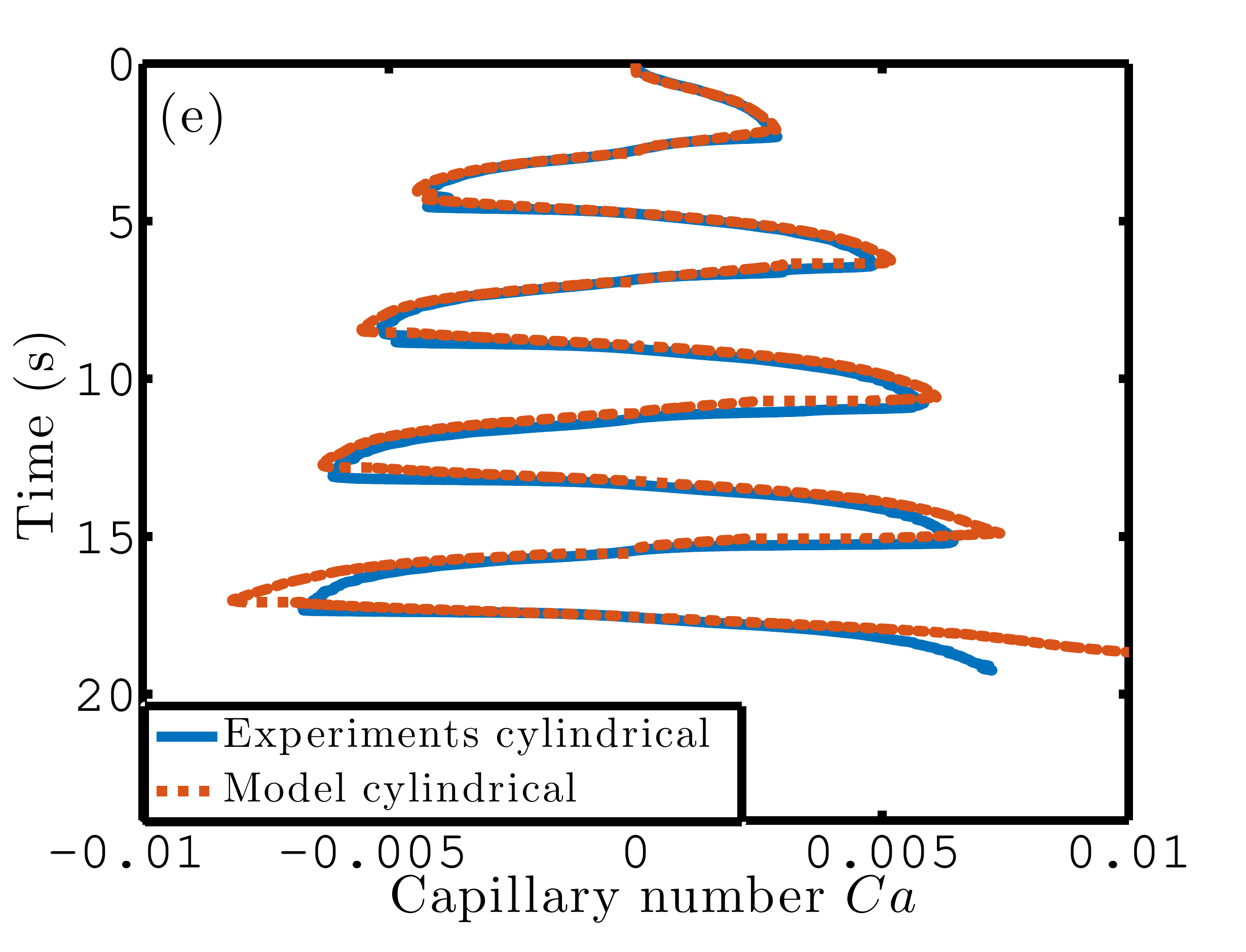}
	\end{subfigure}
	\\
	\centering
	\begin{subfigure}[b]{0.5\textwidth}
		\includegraphics[width=6cm, height=5cm]{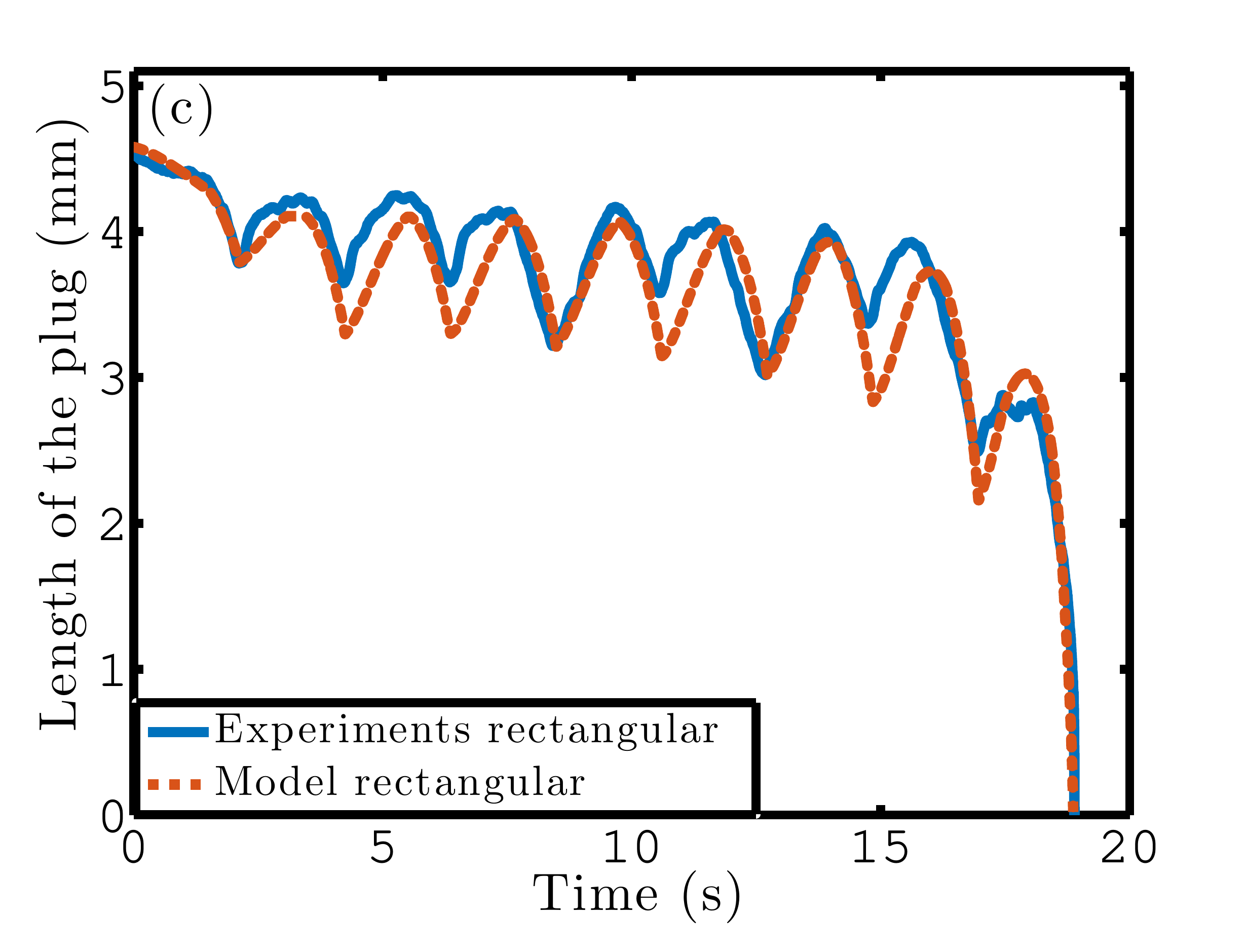}
	\end{subfigure}%
	\begin{subfigure}[b]{0.5\textwidth} 
		\includegraphics[width=6cm, height=5cm]{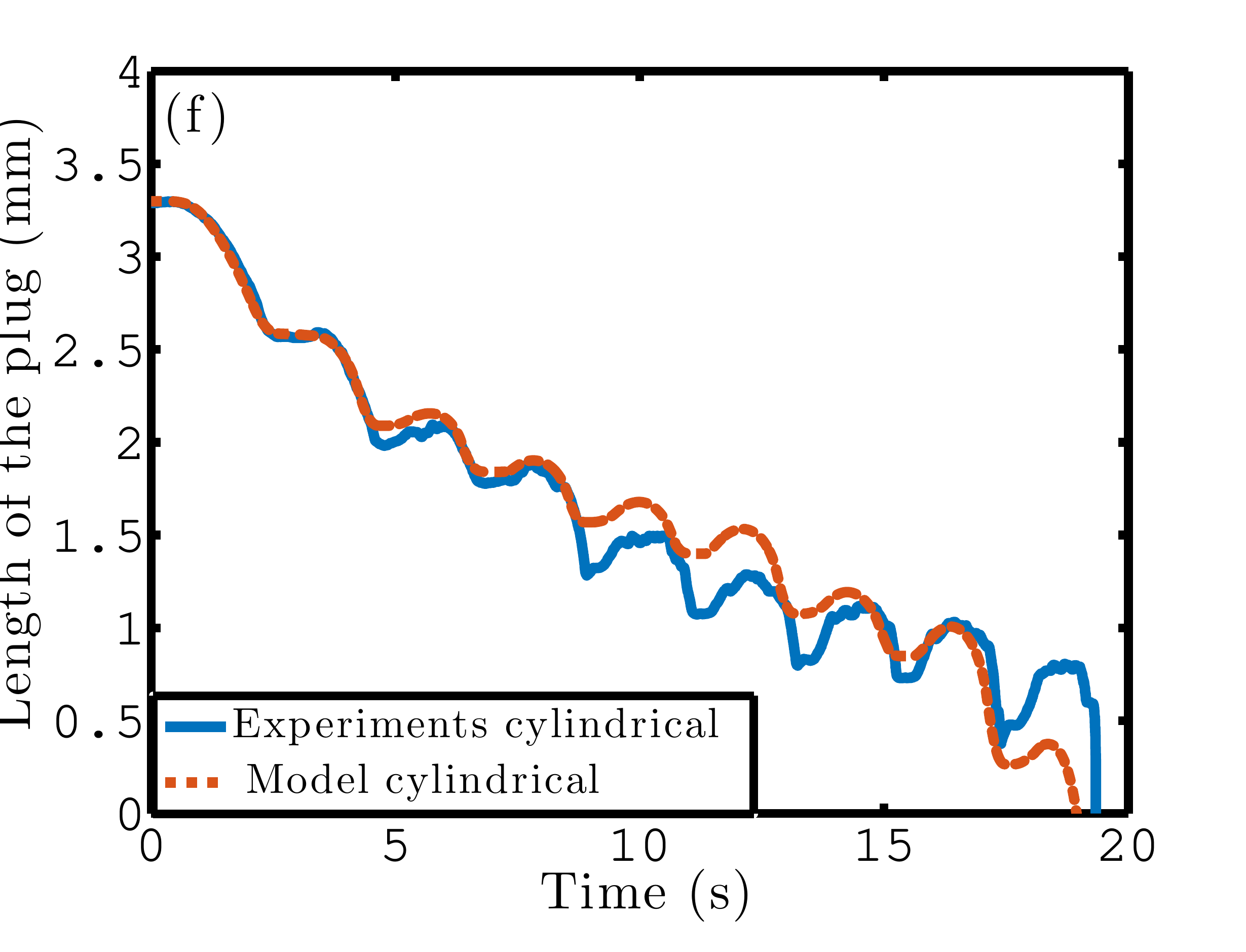}
	\end{subfigure}
	
	\caption{ \label{Cycle_dyn_comparision} (a-b-c) Spatiotemporal evolution of a liquid plug of initial length $L_1 = 4.5 $ mm  moving in a \textit{rectangular microfluidic channel} under the driving pressure represented on \cref{Pressure_forcing}b. (a) Position of the left and right meniscus. (b) Evolution of the capillary number. (c) Evolution of the plug length. (d-e-f) Spatiotemporal evolution of a liquid plug of initial length $L_2 = 3.35$ mm moving in a \textit{cylindrical capillary tube} of diameter $D = 470 \mu$m driven by a cyclic forcing:  $\Delta P_t = 78 e^{-6 e^{-3t}}$ Pa for $t \in [0,T]$, $\Delta P_t = (-1)^n (P_c - P_d)$ for $t \in [nT,(n+1)T]$ with $P_c= 78 e^{-3 e^{-3 (t-nT)}}$ Pa and $P_d = 78 e^{-1.4 (t-nT)} e^{-0.02  e^{-1.4*(t-nT)}}$ Pa, $T = 2.15$ s  with $2T=4$ s. (d) Position of the left and right meniscus. (e) Evolution of the capillary number. (f) Evolution of the plug length. For all experiments, the blue curves correspond to experiments and the red curves to simulations with the model presented in this paper for the experiments (a-b-c) in a rectangular tube and the model presented in \citep{jfm_baudoin_2018} for the experiments (d-e-f) in a cylindrical tube.
	}  
\end{figure}

Such transition is not observed in cylindrical tubes wherein the net variation of the plug size is more regular (see \cref{Cycle_dyn_comparision}f). \cite{jfm_baudoin_2018} demonstrated that in cylindrical channels, the two sources of the plug instability leading to its rupture are (i) the cyclic diminution of the plug viscous resistance to motion due to the diminution of its length and (ii) a cyclic reduction of the plug interfacial resistance due to the deposition of a liquid film of increasing thickness at each cycle and lubrication effects. A very interesting point is that these two instability sources rely on the amount of liquid deposited on the walls. If the amount of liquid left on the walls behind the liquid plug would remain constant, there would be no cyclic evolution of the plug size and no instability related to lubrication effects. Thus the plug would undergo stable periodic motion with no remarkable evolution of its size and no rupture. 

\begin{figure} 
	\centering
	\begin{subfigure}[b]{0.5\textwidth}
	\includegraphics[width=\linewidth , height=5cm]{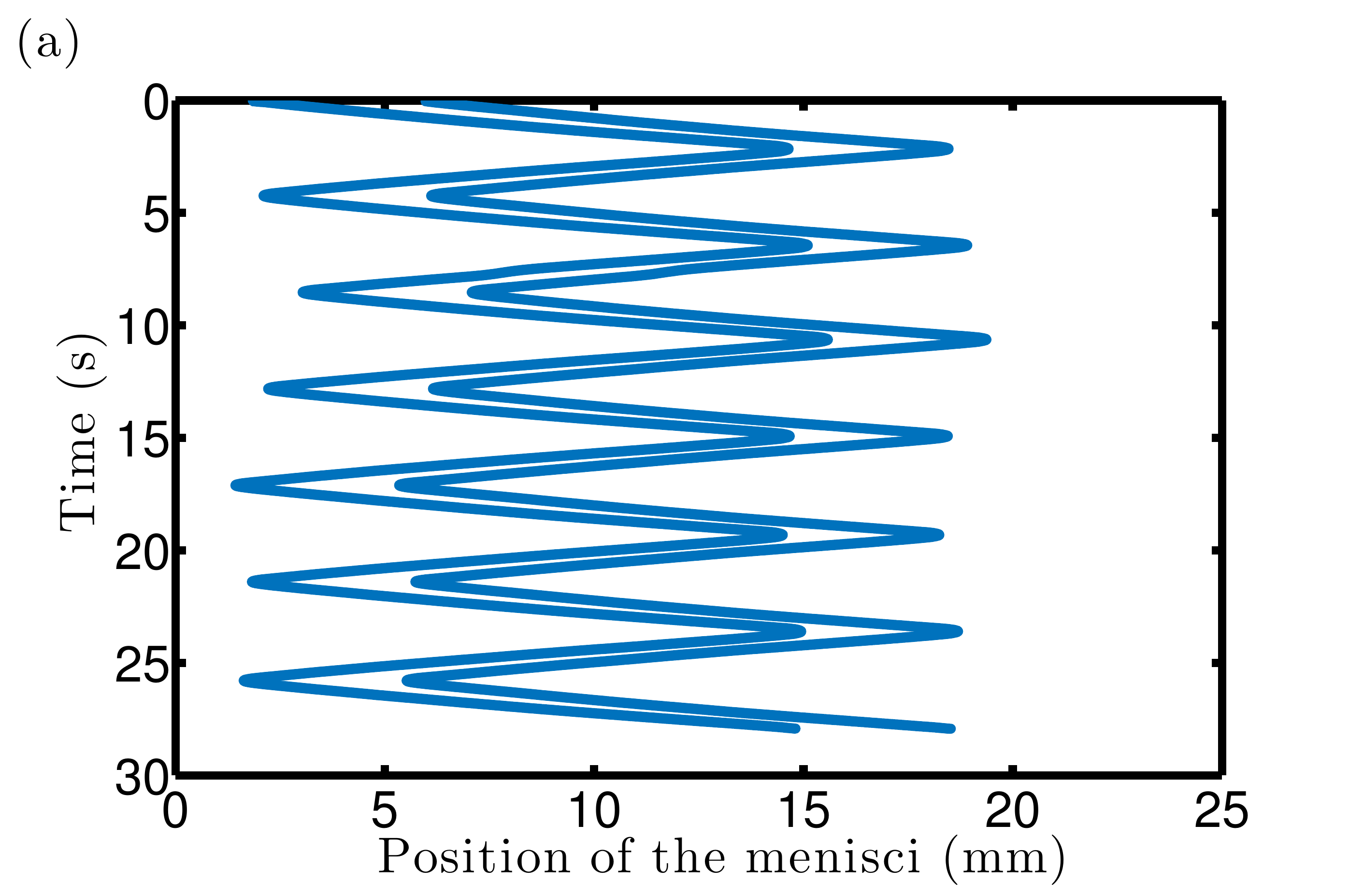}
	\end{subfigure}%
	\begin{subfigure}[b]{0.5\textwidth}
	\includegraphics[width=\linewidth , height=5cm]{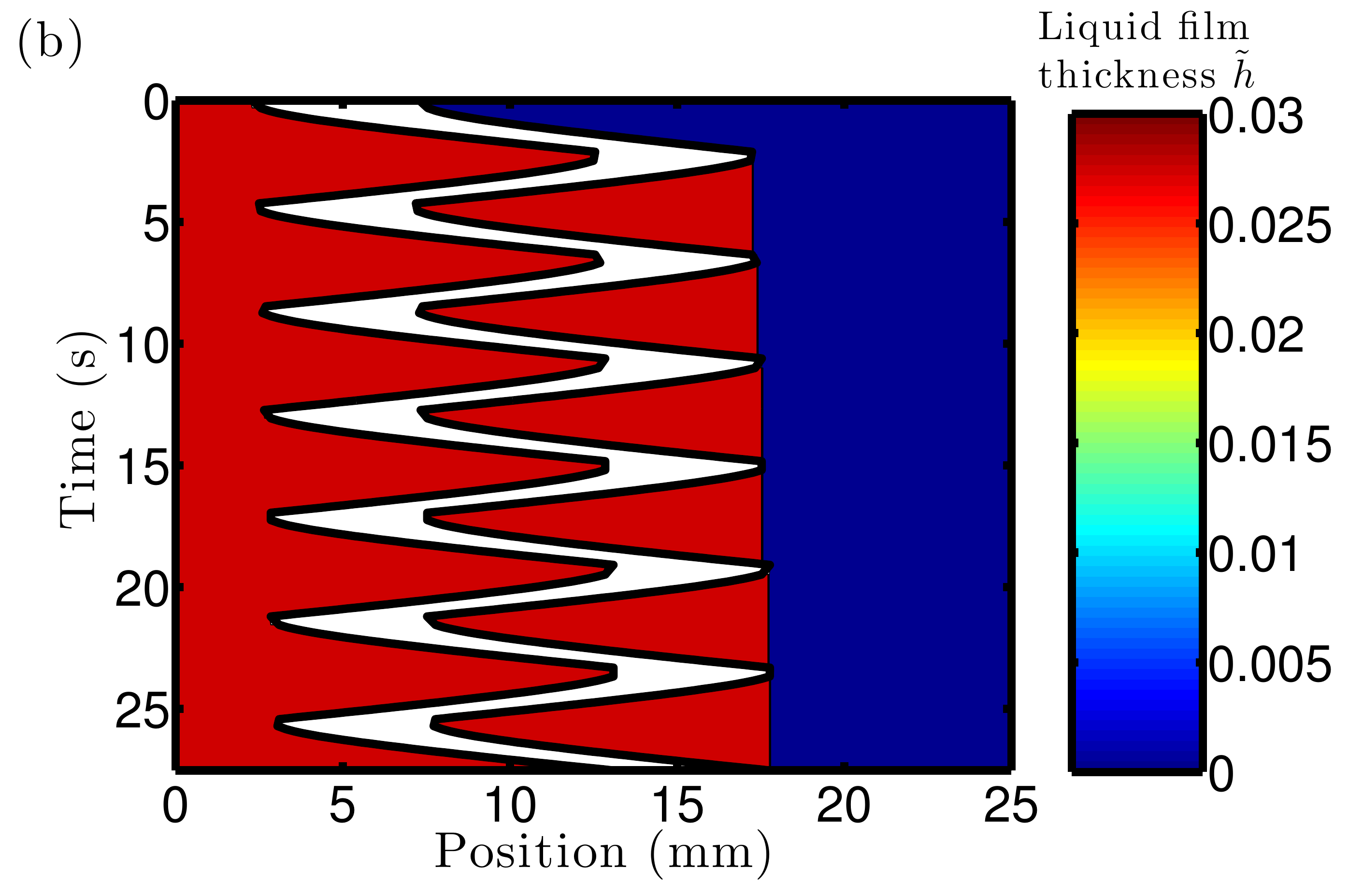}
	\end{subfigure}%
	\caption{ \label{below_CaC} (a) Evolution of the positions of the left and right menisci of a liquid plug of initial length $L_4 = 5$ mm driven by the cyclic forcing represented on \cref{Pressure_forcing}b. We stopped the acquisition after 6 cycles since no significant evolution of the plug length and speed from one cycle to the next was observed. (b) Simulations showing the predicted spatiotemporal evolution of the amount of liquid lying on the walls (wet fraction).
	}  
\end{figure}

This behaviour is indeed observed for plugs of initial length larger than $L_o^c = 4.7$ mm (see \cref{below_CaC}a). In this case the plug always moves at a dimensionless speed smaller than the critical capillary number $Ca_c$.  The simulations are able (i) to quantitatively reproduce the statistical trends of the evolution of the rupture length and rupture time ($80 \%$ of the experimental data match with the simulations within the error bar on the determination of the initial plug length, see Fig. \ref{NormalCascade_setofdatas}) and (ii) to qualitatively reproduce individual dynamics of liquid plugs (see Fig. \ref{Cycle_dyn_comparision}). Thus, we use them below to analyse the evolution of the amount of liquid covering the walls (wet fraction) as a function of time. On \cref{below_CaC}b, we indeed see that the plug leaves a film of constant thickness (constant wet fraction $m_s$), thus leading to a zero cyclic mass balance.

\begin{figure}[h!]
	\centering
	\includegraphics[width=0.8\linewidth]{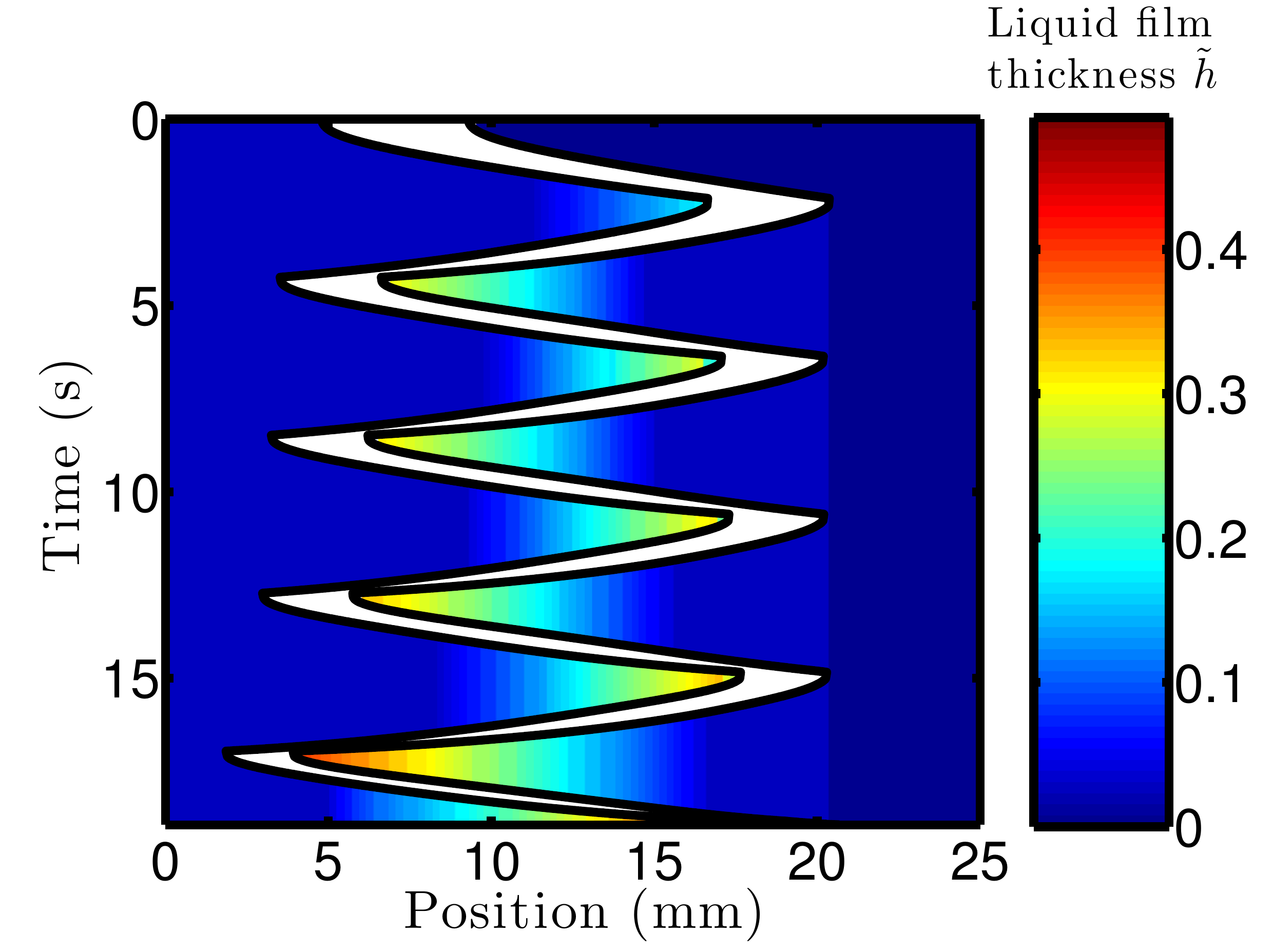}
	\caption{ \label{Spatiotemporal_wetfraction}  Spatiotemporal evolution of the wet fraction for a liquid plug of initial length $L_3 = 4.5$ mm  driven by the cyclic pressure forcing \cref{Pressure_forcing}(b). This figure correspond to the same experiment as \cref{Spatiotemporal_cyclic_motion} (e-f) and  \cref{Cycle_dyn_comparision} (a-b-c)}.
\end{figure}

To understand the transition occurring for liquid plugs smaller than $L_o^c$ we also plotted the evolution of the wet fraction as a function of time for the initial size $L_3 = 4.5$ mm (see \cref{Spatiotemporal_wetfraction}). In this case, the plug moves initially  at a capillary number lying under the critical capillary number $Ca_c$ thus leading to the deposition of a liquid film of constant thickness behind the plug. The plug speed increases  progressively (see \cref{Cycle_dyn_comparision}b) due to the increase in the driving pressure (see \cref{Pressure_forcing}). At time $t \approx 1.5$ s (see \cref{Spatiotemporal_wetfraction}) the plug dimensionless speed overcomes $Ca_c$ and the wet fractions starts increasing until the direction of motion changes (\cref{Spatiotemporal_wetfraction}). During the next cycles, the same behaviour is observed with, at first, the deposition of a film of constant thickness and then the deposition of a film of increasing thickness  (\cref{Spatiotemporal_wetfraction}). Nevertheless, at each cycle, (i) the plug travels further away,  (ii) the plug size decreases, (iii) more and more liquid is left on the walls and (iv) the proportion of the motion above $Ca_c$ increases. For time $t>15$ s  the plug dimensionless speed exceeds $Ca_c$ for the most part of the motion and this leads to a rapid evolution of the plug size and speed and eventually its rupture. This second phase is similar to the evolution of liquid plugs in cylindrical tubes. 

This analysis enables to set a criterion on the stability of a liquid plug driven by a pressure periodic cyclic forcing in a dry rectangular microchannel: If the plug dimensionless speed remains below $Ca_c$ during the first cycle, then the plug dynamics will remain stable during the next cycles, while if the plug reaches $Ca_c$ during this first cycle, it will accelerate cyclically and eventually rupture.

\subsection{Evolution of the rupture time and rupture length and comparison between cyclic and unidirectional forcing.}

\begin{figure}[h!]
	\centering
	\begin{subfigure}[b]{0.5\textwidth}
		\includegraphics[width=6cm, height=5cm]{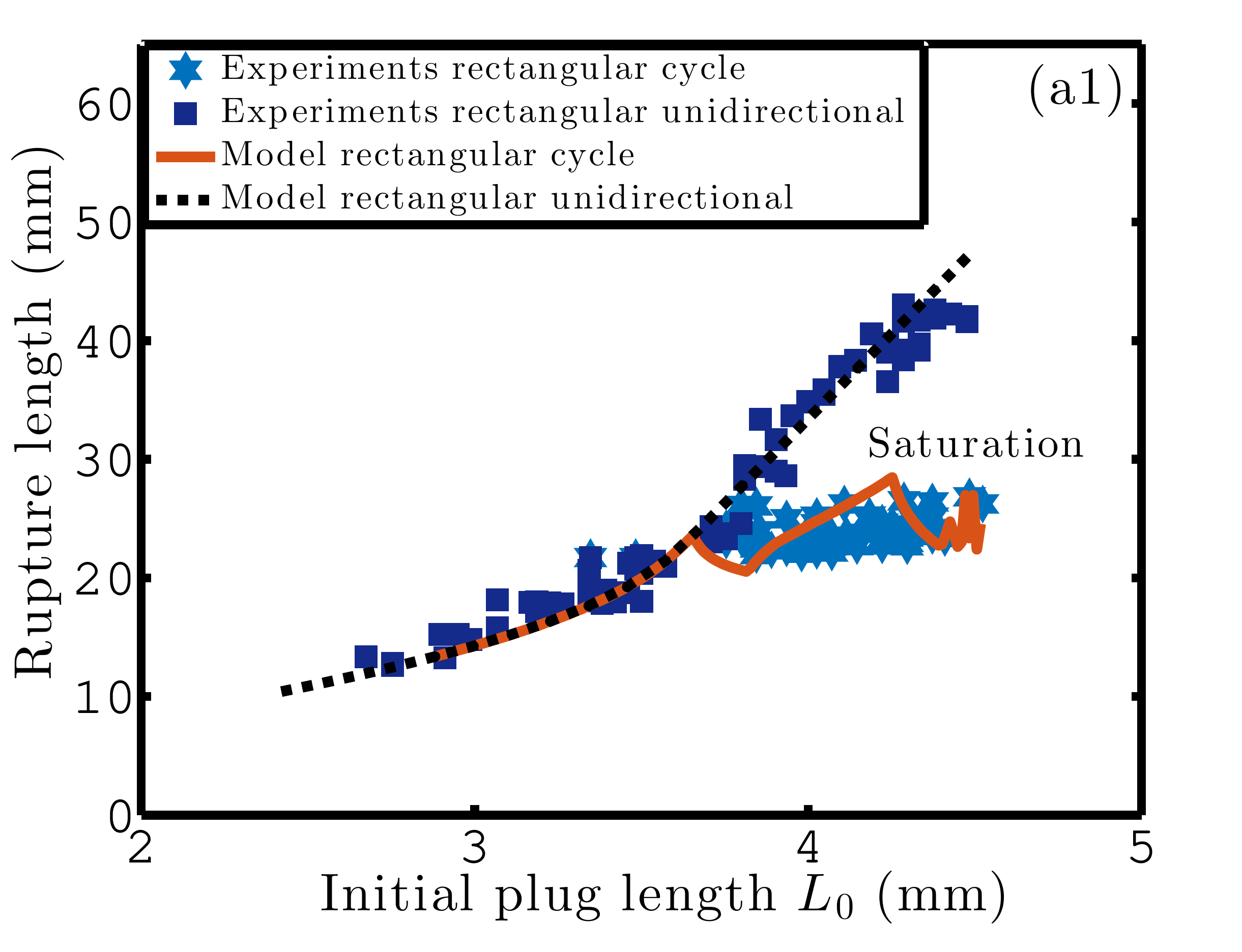}
	\end{subfigure}%
	\begin{subfigure}[b]{0.5\textwidth}
		\includegraphics[width=6cm, height=5cm]{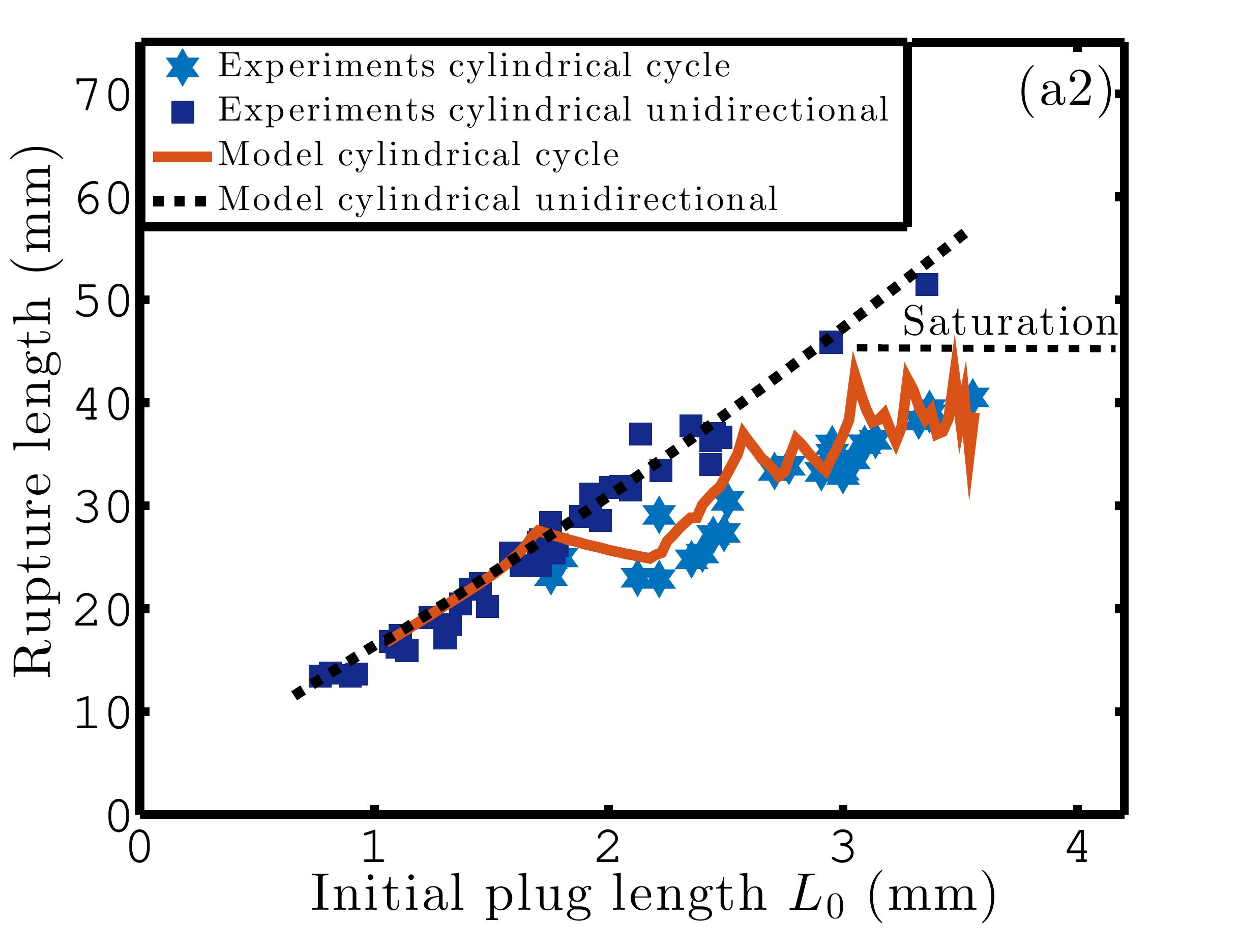}
	\end{subfigure}%
	\\
	\centering
	\begin{subfigure}[b]{0.5\textwidth}
		\includegraphics[width=6cm, height=5cm]{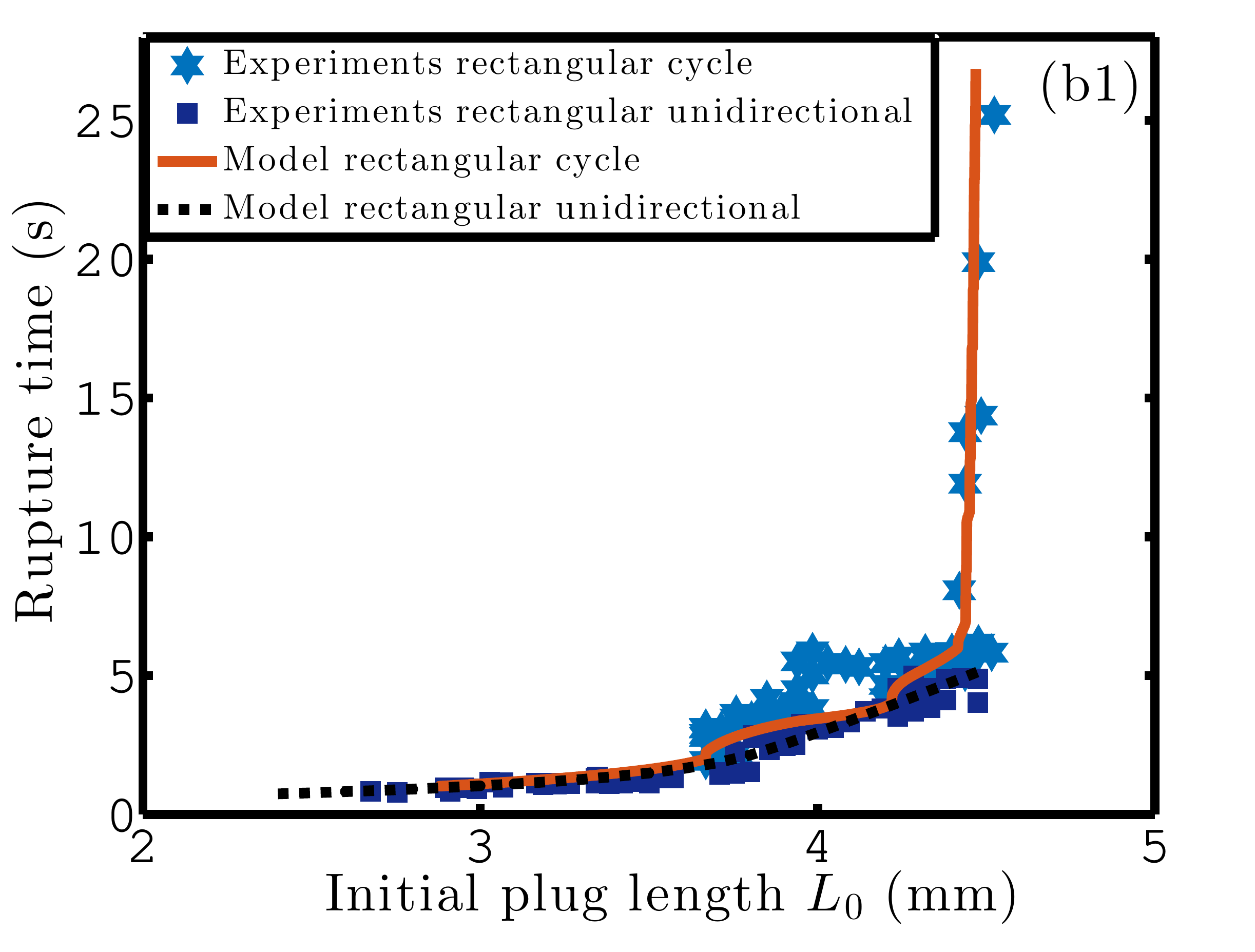}
	\end{subfigure}%
	\begin{subfigure}[b]{0.5\textwidth}
		\includegraphics[width=6cm, height=5cm]{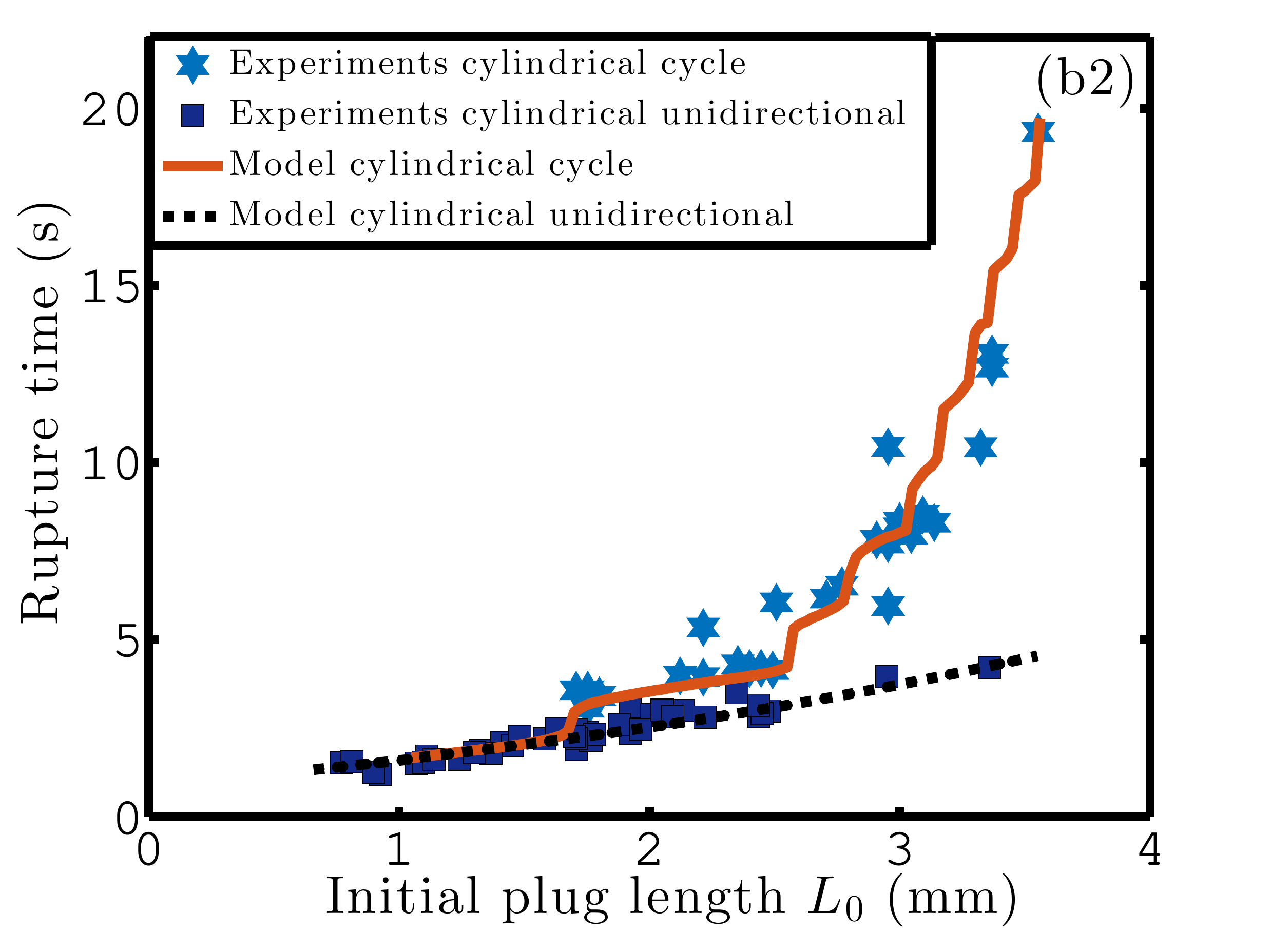}
	\end{subfigure}
	
	\caption{ \label{Setofdatasincycle_compare_Cy_Rec}  
		(a1-a2) Rupture lengths  and  (b1-b2) rupture times for a cyclic or unidirectional pressure forcing in a rectangular (a1-b1) or cylindrical (a2-b2) channel. For rectangular channels, the pressure driving corresponds to the one represented on \cref{Pressure_forcing} and for the cyclic pressure driving, it correspond to the one described in \cref{Cycle_dyn_comparision}. Blue stars correspond to experiments with the cyclic forcing, black squares to experiments with the unidirectional pressure forcing, red solid lines to simulations for the cyclic pressure forcing and black dashed lines to simulations with the unidirectional pressure driving. }  
\end{figure}

To get a parametric overview of the liquid plugs dynamics in rectangular microchannels, we performed hundreds experiments with different plug initial lengths and either unidirectional or cyclic pressure drivings (of same maximal amplitude). The measured values of the rupture time and rupture length are represented on \cref{Setofdatasincycle_compare_Cy_Rec} (a1-b1) and compared to the evolutions in cylindrical tubes (a2-b2). As previously reported in cylindrical channels, we observe a saturation of the rupture length when the plug starts undergoing cycles. Nevertheless a major difference with cylindrical tubes is that the rupture time increases to infinity for a finite value of the critical initial length $L_o^c \approx 4.7$ mm while the increase in the rupture time was shown to follow a more "gradual" exponential trend in \cite{jfm_baudoin_2018}. This is again a consequence of the existence of the quasi-static deposition regime in rectangular channels which does not exist in cylindrical tubes. Of course, in any case, the rupture time is never really infinite owing to evaporation of the plug occurring in the channel and the weak dependence over the capillary number.

\section{Critical assessment of the experimental dispersion and the model validity.}

The experiments presented in this paper were extremely sensitive on the contamination of the channels. To avoid pollution, (i) the channels were fabricated in the clean room and stored in a  sealed box until their use, (ii) the air injected in the channels by the pressure controller was filtered by several filters, and (iii) we were extremely cautious in the liquid sampling to avoid pollution of Perfluorodecalin. Nevertheless, even with all these precautions, we had to change the channels regularly when contamination was observed to avoid deviation in the results. We believe that this pollution is the main source of discrepancy in the experiments. From the theoretical side, the main shortcomings of the model are:

\begin{enumerate}
\item The approximation of the progressive transition between the quasi-static and dynamic deposition regimes (observed in \cite{de2007scaling}) by a sharp transition between a constant value behind a critical capillary number and equation \eqref{wet_frc} above.
\item The omission of all the unsteady and convective terms from Stokes equation, even in the most accelerative phase and the approximation of the viscous pressure drop by a Poiseuille law.
\item The 2D (high aspect ratio) approximation for the estimation of the front interface pressure drop.
\item The approximation of the tube prewetting film by the formula  $\sqrt{1 - \tilde{S}_f}$.
\end{enumerate}
With that said, it  is difficult to determine the exact origin of discrepancies between experiments and theory. In particular we are unable at the present state to determine whether observed deviations mostly originate from experimental imperfections or approximations in the theory.

\section{Conclusion}
 In this work, we studied the dynamics of single liquid plugs in rectangular microfluidic channels under unidirectional and cyclic pressure forcings. First we showed that the transition between quasi-static and dynamic film deposition regimes leads to a dramatic acceleration of the plug rapidly leading to its rupture. A pressure-dependent critical size for the transition between these two regimes is derived analytically. For cyclic periodic pressure forcing, we showed that two regimes can occur depending on the initial size of the plug: the plug can either undergo stable periodic oscillations or cyclically accelerate and eventually rupture. The stable regime is observed when the plug dimensionless speed remains below a critical capillary number during the first cycle, while the second is observed as soon as the plug overcomes this value during the first cycle. We were able to quantitatively reproduce the evolution with a reduced dimension model obtained from the combination of previous elements introduced by  \cite{baudoin2013airway}, \cite{magniez2016dynamics} and \cite{jfm_baudoin_2018} with additional elements to consider the transition between the quasi-static and dynamic film deposition regimes. 

These results are of primary interest since microfluidic channels with rectangular cross sections are widely used in the field of microfluidics owing to their easy fabrication. In particular, for the study of liquid plugs dynamics in complex geometries, such as airway tree, it is extremely difficult to design trees with cylindrical sections. Thus this work also enables to analyse and transpose results obtained in rectangular channels to cylindrical channels and understand the pertinence and limit of such comparison.

\section{Acknowledgment}
We acknowledge stimulating discussions with J.C. Magniez. This work was supported by Universit\'{e} de Lille.

\bibliographystyle{model2-names.bst}\biboptions{authoryear}
\bibliography{Bibliography_Article,Bibliography_jfm}

\end{document}